\documentclass[aps, prb, reprint, longbibliography, floatfix, superscriptaddress, 10pt]{revtex4-2}
\usepackage[utf8]{inputenc}
\usepackage{amsmath}
\usepackage{amssymb}
\usepackage{braket}
\usepackage[american]{babel}
\usepackage{csquotes}
\MakeOuterQuote{"}
\usepackage{graphicx}
\usepackage{xcolor}
\usepackage[caption=false]{subfig}
\usepackage{hyperref}
\hypersetup{colorlinks, citecolor=blue, linkcolor=violet, urlcolor=blue}

\newcommand{\pauli}[3][\sigma]{#1^{#2}_{#3}}

\newcommand{\opnamed}[2]{#1_{\mathrm{#2}}}

\newcommand{\irm}{\mathrm{i}}
\newcommand{\erm}{\mathrm{e}}
\newcommand{\drm}{\mathrm{d}}

\DeclareMathOperator{\sign}{\mathrm{sign}}
\DeclareMathOperator{\Tr}{\mathrm{Tr}}

\graphicspath{{Figs/}}
\begin{document}
\title{Diabatic quantum and classical annealing of the Sherrington-Kirkpatrick model}

\author{Artem Rakcheev}
\email{artem.rakcheev@psi.ch}
\affiliation{Laboratory for Theoretical and Computational Physics, Paul Scherrer Institut, 5232 Villigen PSI, Switzerland}

\author{Andreas M. Läuchli}
\affiliation{Laboratory for Theoretical and Computational Physics, Paul Scherrer Institut, 5232 Villigen PSI, Switzerland}
\affiliation{Institute of Physics, Ecole Polytechnique Fédérale de Lausanne (EPFL), CH-1015 Lausanne, Switzerland}

\date{\today}

\begin{abstract}
Quantum annealing is a contender to solve combinatorial optimization problems based on quantum dynamics. While significant efforts have been undertaken to investigate the quality of the solutions and the required runtimes, much less attention has been paid to understanding the dynamics of quantum annealing and the process leading to the solution during the sweep itself. In this comprehensive study, we investigate various aspects of the quantum annealing dynamics using different approaches. We perform quantum annealing, simulated quantum annealing, and classical annealing on several hundred instances of the Sherrington-Kirkpatrick model with intermediate system sizes up to $22$ spins using numerical simulations. We observe qualitative differences between the quantum and classical methods, in particular at intermediate times, where a peak in the fidelity, also known as diabatic bump, appears for hard instances. Furthermore, we investigate the two-point correlation functions, which feature differences at intermediate times as well. At short times, however, the methods are similar again, which can be explained by relating the short-time expansion of quantum annealing to a high-temperature expansion, thus allowing in principle to find the classical solution already at short times, albeit at prohibitive sampling cost.  
\end{abstract}
\maketitle

\tableofcontents

\section{\label{sec: intro} Introduction}
In 1998, it was suggested by Kadowaki and Nishimori to use quantum fluctuations to solve combinatorial optimization problems~\cite{kadowaki_quantum_1998}. To be precise, a noncommuting "driver" Hamiltonian is added to a problem Hamiltonian, which is diagonal in the computational basis and whose ground state encodes the solution of the optimization problem. Starting in the ground state of the driver, which is very strong at this point, the strength is reduced dynamically, and the state evolves according to the Schrödinger equation, hopefully having a large overlap with the desired ground state. The resulting method, named quantum annealing by its inventors, received a lot of attention in the following decades~\cite{albash_adiabatic_2018, hauke_perspectives_2020, rajak_quantum_2022} as it might outperform classical algorithms for relevant applications and as experimental realizations using superconducting qubits exist and can operate on up to several thousand qubits~\cite{cohen_dwave_2014, boixo_evidence_2014, denchev_what_2016, king_quantum_2023}. Most of the early studies focused on long annealing times, where the method can be related to the adiabatic theorem, guaranteeing to find the (desired) ground state if the fluctuations are weakened adiabatically i.e. slowly enough~\cite{morita_mathematical_2008, farhi_quantum_2000, bachmann_adiabatic_2017}. At long annealing times, quantum annealing, therefore, corresponds to adiabatic quantum computation. 

Of course, the main question here is, how long these times need to be for a given problem and how they compare with classical algorithms. The decisive quantity for quantum annealing is the minimal gap between the instantaneous ground state and excited state. This gap can be related to the order of the quantum phase transition, with typically polynomially small gaps in the system size at second order and exponential small gaps at first order transitions. However, some works show that at first order transitions the situation can be much more varied~\cite{laumann_quantum_2012, tsuda_energy_2013, laumann_quantum_2015}. The next question of how this compares with classical algorithms is also strongly model-dependent. Here, mixed results were reported in the literature for different problems, with some problems benefiting from quantum annealing~\cite{farhi_quantum_2002}, while other even very simple problems, such as the mean-field all-to-all ferromagnet, lack such benefits and can or even show worse performance~\cite{jorg_energy_2010, bapst_quantum_2012, liu_quantum_2015, heim_quantum_2015}. Furthermore, many works suggest that quantum annealing of NP-hard problems is exponentially hard in the quantum case as well~\cite{young_size_2008, altshuler_anderson_2010, hen_performance_2015, hen_exponential_2011, knysh_zerotemperature_2016} or at least that an exponential speedup can not be expected for all types of problem instances~\cite{katzgraber_seeking_2015}. However, some works do report positive results for different models~\cite{ king_scaling_2021, yan_analytical_2022, king_quantum_2023}. Furthermore, non-exponential speedups could also be possible and useful. Even given a benefit, a major question is, whether the actual experimental annealers could use it since they most likely do not implement the coherent unitary dynamics of the Schrödinger equation, but suffer from finite-temperature and other imperfections~\cite{cohen_dwave_2014, hauke_perspectives_2020}. In any case, interest in quantum annealing with faster annealing times outside the adiabatic regime grew recently, and this setup is also known as diabatic quantum annealing~\cite{crosson_prospects_2021}. However, to our best knowledge, there are few works that have investigated quantum annealing for various annealing times~\cite{katsuda_nonadiabatic_2013, callison_energetic_2021}.

In this article, we present a study comparing quantum annealing and two classical annealing algorithms, simulated annealing and simulated quantum annealing on the Sherrington-Kirkpatrick model (mean-field spin glass); a problem known to be NP-hard problem~\cite{barahona_computational_1982}. We investigate two to three hundred instances for system sizes ranging from $N=8$ to $N=22$ and perform the different annealing types at various speeds. This allows us to gain an overview of the qualitative behavior of the methods at short, intermediate, and long times. We will see that the methods show similarities in the short and long time regimes, and that most qualitative deviations between the methods lie in the intermediate time regime. In fact, for our model and instances, the qualitative deviations are strong between quantum annealing and the classical annealing methods, with both classical methods being rather similar. The article is structured as follows: in Sec.~\ref{sec: annealing methods} we introduce the different annealing methods in more detail and comment on the numerical implementations. Next, in Sec.~\ref{sec: annealing model}, we introduce the model and discuss the instances which were investigated. We then discuss the behavior of the fidelity and the correlation functions in Sec.~\ref{sec: annealing fidelity} and~\ref{sec: annealing corrs}. Here, we focus on identifying the different regimes discussed above and on identifying the differences and similarities between the methods. It will turn out, that particularly at short times the methods are very similar to each other in terms of the correlation functions and fidelity growth. We argue that this is related to the probability distribution in the computational basis in Sec.~\ref{sec: annealing dist}. In particular, we demonstrate that at short times quantum annealing produces a thermal distribution at high temperatures.

Throughout the article, we use dimensionless units obtained by setting $\hbar=1$ and $k_{B}=1$. Furthermore, we measure energies in units of a fictitious interaction constant $J$, which we also set to unity. In practice, this value would be determined by the strength of the interactions in the Sherrington-Kirkpatrick Hamiltonian (Eq.~\eqref{eq: SK model}).  
\section{\label{sec: annealing methods}Annealing methods}
\subsection{\label{ssec: annealing methods - qa}Quantum annealing}
Quantum annealing (QA) works by evolving a state with a time-dependent Hamiltonian interpolating between an initial and final Hamiltonian $\opnamed{H}{ini}$ and $\opnamed{H}{fin}$
\begin{equation}
\label{eq: sweep ham def}
H(s(t))=(1-s(t))\opnamed{H}{ini}+s(t)\opnamed{H}{fin},
\end{equation}  
parameterized by a time-dependent parameter $s$ running from $0$ to $1$, while $t$ runs from $0$ to the annealing time $T$. In this work, we will focus on a standard setup with $s(t)=t/T$. The initial state is the ground state for $\opnamed{H}{ini}$. A typical choice, which we will also use, is to take a transverse field $\opnamed{H}{ini}=\opnamed{H}{x}=-\sum\limits_{i}\pauli{x}{i}$ for quantum fluctuations. As a result, the model which we treat is not the most generic version of quantum annealing, since the total Hamiltonian is a so-called \emph{stoquastic} Hamiltonian i.e. all off-diagonal matrix elements are non-negative~\cite{albash_adiabatic_2018}. This property also enables simulated quantum annealing, which we will introduce later in this section. The differences of stoquastic and general Hamiltonians for quantum evolution are a subject of current research~\cite{crosson_designing_2020, crosson_rapid_2021, hastings_power_2021}.

We simulate quantum annealing numerically, by discretizing time with a resolution of $\Delta t = 0.01$, and then approximating the time-dependent Hamiltonian by a fourth-order commutator-free Magnus expansion~\cite{alvermann_highorder_2011,blanes_magnus_2009}. This results in having to compute the action of exponentials of large sparse matrices onto the state at each step. This can be done efficiently using Krylov subspace methods~\cite{higham_functions_2008, arbenz_solving_2016,saad_iterative_2003, saad_numerical_2011, lauchli_numerical_2011} with partial reorthogonalization~\cite{simon_lanczos_1984} and an appropriate error bound~\cite{wang_error_2017} to ensure the accuracy of the result. The model which we will use has a "spin-flip" symmetry, meaning that one can exchange $\pauli{z}{i}$ by $-\pauli{z}{i}$ at all sites $i$, without changing the Hamiltonian. As a result, the full Hilbert space can be decomposed into two independent subspaces with even and odd spin-flip parity. Since the initial state lies fully in the positive parity subspace, we can operate therein and ignore the other. This reduces the Hilbert space dimension by a factor of two, thus the size of the dimension for $N$ spins is $2^{N - 1}$. For the largest system size of $N=22$, this amounts to a dimension of $2097152$. During the simulations, we compute the two-site correlation functions $ G_{ij}=\braket{\pauli{z}{i} \pauli{z}{j}} $ and record the $1000$ most probable states in the computational basis, to gain an overview of the distribution.
\subsection{\label{ssec: annealing methods - sa}Simulated annealing}
Simulated annealing (SA) is the classical "predecessor" to quantum annealing. It mimics the cooling down of a classical system, which, if done slowly enough, results in the system reaching the equilibrium state at zero temperature; the ground state of the system~\cite{kirkpatrick_optimization_1983, zomaya_simulated_2010}. Thermalization dynamics with a time-dependent temperature and associated inverse temperature $\beta$, can be described by the classical master equation
\begin{equation}
\label{eq: master eq.}
\dot{P}(t)=Q(\beta(t))P(t),
\end{equation}
where $P$ denotes the probabilities of the classical states and $Q$ is the so-called \emph{transition rate matrix}. This equation is extremely similar to the Schrödinger equation and the state at time $t$ can be obtained from the \emph{transition matrix} 
\begin{equation}
\label{eq: transition matrix}
W(t)=\mathcal{T} \exp \left(\int \limits_{0}^{t} Q(\beta(\tau)) \; \drm \tau \right),
\end{equation}
by $P(t)=W(t)P(0)$. If the temperature is constant, this describes equilibration at a constant temperature, but if it is lowered from infinite (or very high) to zero (or very low) temperature sufficiently slowly the state remains in equilibrium, and the system is annealed to the ground state.

Here, one should note the similarity with the quantum case, where one can define the time-evolution operator
\begin{equation*}
    U(t)=\mathcal{T} \exp \left(-\irm \int \limits_{0}^{t} H(\tau) \; \drm \tau \right)
\end{equation*}
analogously. Of course the mathematical properties of $Q$ and $W$ have to be different than the ones of $H$ and $U$ in the quantum case, since the probabilities satisfy $\sum\limits_{n}P_{n}=1$ contrary to the amplitudes $a_n$ satisfying $\sum\limits_{n} |a_{n}|^{2}=1$ in the quantum case (in both cases $n$ refers to states in some arbitrary basis). As such a matrix like $W$, conserving the sum of probabilities, is also called a \emph{stochastic matrix}~\cite{ norris_markov_1997}.

Here, we do not want to focus on these details though, since one of the main advantages of SA is that one can avoid simulating the master equation using the full probability vector and transition rate matrix, which would be exponentially costly as in the case of QA. Rather, one can simulate the equation stochastically, keeping only one state (or a relatively small number of states) at a time and never using the full matrix. Usually a \emph{Markov chain Monte Carlo} approach is taken here~\cite{santoro_theory_2002, battaglia_deterministic_2005, heim_quantum_2015}. For this, one first assumes a discretization in time, with the state of the system being one of the possible classical states i.e. a state in the $z$-basis. In each time step, also called a Monte Carlo step (MCS), the state is either changed or stays constant. This change is non-deterministic, but depends on a random number, leading to the aforementioned stochastic simulation. If the transition rates $Q_{ij}$ \emph{or} the transition probabilities $W_{ij}$ from state $i$ to state $j$ satisfy \emph{detailed balance}~\cite{norris_markov_1997, landau_guide_2014}
\begin{equation}
\label{eq: detailed balance}
W_{ji}\rho_{i}=W_{ij}\rho_{j} \; \forall i\neq j
\end{equation}
with the Boltzmann weights $\rho_{i}=\exp(-\beta E_{i})/Z(\beta)$, averaging over several simulations yields the solution of the classical master equation~\footnote{It may seem surprising, that the detailed balance is the same for $Q$ and $W$, however, they do differ in the diagonal entries.}. Here, the energies $E_{i}$ are the eigenvalues of the (classical) Hamiltonian, and it is important to note that the detailed balance condition in Eq.~\eqref{eq: detailed balance} is reversed in terms of the indices compared to most literature. The reason is that historically Eq.~\eqref{eq: master eq.} is typically written as $P(t)Q$, with $Q$ acting \emph{to the left} and the same applies to $W$. Here we choose a different convention to highlight the similarities to the Schrödinger equation.

For our simulations, we use a simple scheme based on \emph{single spin-flips}, meaning that at each step a new state, differing from the current one by a single randomly flipped spin is proposed, and the acceptance probability is given by the \emph{Glauber rule}
\begin{equation}
\label{eq: glauber rule}
P_{\mathrm{acc}}(\beta,\Delta E)=\frac{1}{2} \left[ 1- \tanh \left( \frac{\beta \Delta E}{2} \right) \right],
\end{equation}
with $\Delta E$ the energy difference between the proposed and the current state. The Glauber rule is known to be very similar to the Metropolis-Hastings rule, but can have advantages at high temperatures~\cite{landau_guide_2014}. For the model defined in Eq.~\eqref{eq: SK model}, the energy difference upon flipping spin $i$ is 
\begin{equation*}
\Delta E = -\sigma^{z}_{i}\sum\limits_{j}J_{ij}\pauli{z}{j}.
\end{equation*} 
The annealing time is set by the number of MCS, where the value of $s$ changes linearly between $0$ and $1$ during the entire run and the temperature is given by $\beta(s)=(1/s-1)^{-1}$. In that sense, having more steps leads to a slower temperature variation and gives the system more time to potentially equilibrate in a temperature window. The temperature schedule is derived by equating the temperature $T$ with the ratio of the transverse field strength to the Ising strength $\Gamma$ following~\cite{kadowaki_quantum_1998}. In our case, the ratio as a function of $s$ is $(1-s) / s$. We typically perform simulations with $100-10000$ MCS for each instance with a resolution of $ \Delta \mathrm{MCS} = 100$. As we will see in later sections, this is typically enough to reach the long-time limit. Averaging over $R$ runs at a fixed value of MCS, yields expectation values for the correlation functions, as well as the distribution of the most probable $1000$ states for each value of $s$. These values are expected to converge with $1 / \sqrt{R}$~\cite{landau_guide_2014}, thus we perform $R=10^{6}$ runs, to resolve even relatively small correlations. 

To be precise, we start each run with a random state (state here refers to a product state in the computational basis), such that the average corresponds to infinite temperature, and then keep a record of states after at every measuring step. There are $100$ measuring steps, independent of the number of MCS, such that we get the same resolution in $s$, $\Delta s = 0.01$, as in QA. These records are kept for every run, such that at the end we have a tally, of how often each state appeared at each step over all runs. From this, we can obtain the expectation value of the correlation functions as well as the distribution of the $1000$ most probable states at each step. Of course, we actually have all states and not just $1000$, but saving all states for all instances and annealing times would require far too much storage: a quick estimate for $N=20$ assuming $100$ instances, $100$ simulations per instance, and $100$ measurements per simulation gives $2^{19} \cdot 10^{6} \cdot 4 \; \mathrm{B}\approx 2\; \mathrm{TB}$! 
\subsection{\label{ssec: annealing methods - sqa}Simulated quantum annealing}
Simulated quantum annealing (SQA), sometimes also referred to as quantum simulated annealing or path integral Monte Carlo, is essentially again classical simulated annealing, but this time performed with a Hamiltonian derived from the QA protocol via the quantum-classical correspondence, rather than the solely classical Hamiltonian~\cite{santoro_theory_2002, battaglia_deterministic_2005, heim_quantum_2015}. The main idea is that finite-temperature properties of a $d-\text{dimensional}$ quantum Hamiltonian can be mapped to the finite-temperature properties of a $(d+1)-\text{dimensional}$ classical Hamiltonian. For our finite-size spin systems, this mapping applied to the quantum Hamiltonian with $N$ spins at temperature $\beta$, leads to a classical Hamiltonian with $N \times n$ spins at temperature $\beta_{\mathrm{SQA}}=\beta / n$~\footnote{Note that the couplings in the mapped Hamiltonian may depend on $\beta_{\mathrm{SQA}}$ though, so the interpretation as a temperature may not strictly be appropriate.}. If $\beta$ is chosen sufficiently high, we can now follow the thermal properties close to the ground state of the QA Hamiltonian. Hence, although the unitary evolution of QA is \emph{not} simulated in SQA~\cite{bando_simulated_2021}, the algorithm still uses some features of the quantum Hamiltonian and as such can even be \emph{exponentially} faster than SA~\cite{santoro_theory_2002, bapst_thermal_2013, heim_quantum_2015, crosson_simulated_2016, albash_simulatedquantumannealing_2016}, although the full conditions, for which a speedup occurs are not yet understood in full.

 The quantum-classical mapping relies on relating the partition function $Z(\beta)$ of a suitable quantum model, to the partition function of a classical model. An introduction can be found in many textbooks, for example in~\cite{landau_guide_2014}. For the Sherrington-Kirkpatrick model from Eq.~\eqref{eq: SK model} in a transverse field, the "extended" classical Hamiltonian is
\begin{equation}
\label{eq: mapped ham}
\opnamed{H}{cl}=\frac{s}{2} \sum\limits_{i, j=1}^{N} \sum_{k=1}^{n} J_{ij}z^{(k)}_{i}z^{(k)}_{j} +J_{\perp}(s) \sum\limits_{i=1}^{N}\sum_{k=1}^{n}z^{(k)}_{i}z^{(k+1)}_{i},
\end{equation}
here the upper index is just a dummy index numbering the corresponding insertion. The states numbered in this way are sometimes called \emph{replicas}, and we will use this terminology as well. Periodic boundary conditions are assumed for the inter-replica interaction, whose coupling is ferromagnetic and given by 
\begin{equation*}
    J_{\perp}(s)=\frac{1}{2 \beta_{\mathrm{SQA}}} \ln\left[\coth\left( \beta_{\mathrm{SQA}}(1-s)\right)\right].
    \end{equation*}
In the continuum limit, $n \to \infty$, the ground state of this Hamiltonian will correspond to the thermal state of the quantum model. Varying $s$, and therefore the coupling results in dynamics, which could feature properties of the full quantum dynamics. For the simulations, we use $n=8$ replicas. While this is a relatively small number, which is far from the continuum limit, research suggests~\cite{heim_quantum_2015} that this number leads to a very good performance from an optimization perspective. We again perform SQA based on single spin-flips using the Glauber rule with a fixed number of MCS, this time at a constant $\beta=10$, changing only the coupling $J_{\perp}$. For each simulation, we again use $100$ measurements based on a resolution $\Delta s = 0.01$ and average over $R=10^{6}$ runs. Additionally, we also again record the most probable $1000$ states, this time \emph{per replica}.

\section{\label{sec: annealing model}Model and instances}
\subsection{\label{ssec: annealing model - sk}Sherrington-Kirkpatrick model}
For the optimization problem, we choose the Sherrington-Kirkpatrick model for $N$ spins
\begin{equation}
\label{eq: SK model}
\opnamed{H}{fin}\equiv \opnamed{H}{SK}=\frac{1}{2}\sum\limits_{i, j=1}^{N}J_{ij}\pauli{z}{i}\pauli{z}{j},
\end{equation}
where the $J_{ii} = 0$ and $J_{ij}=J_{ji}=\pm 1$ are random variables with $\pm 1$ being equally likely (bimodal distribution). Another common choice for the bonds $J_{ij}$ is a Gaussian distribution. In fact, there are more analytical results on the latter, however, the former seems simpler, while still being NP-hard~\cite{barahona_computational_1982}, since only the number of (un-)satisfied bonds is relevant without weights. Satisfied bonds are those, where the connected spins are aligned in the low-energy configuration with respect to the bond. Models, where not all bonds can be satisfied simultaneously are often called \emph{frustrated} models, with additional \emph{disorder}, here in the form of random bonds, often leads to spin glass behavior~\cite{fischer_spin_1991}. While spin glasses can feature fascinating and exotic behavior, due to a complex energy landscape, a detailed discussion of spin glass physics is beyond the scope of this article, and the observations we will make and the arguments we will develop, should not depend on the general properties on the model. We will rather focus on properties of specific instances, which may well be present in instances of other Ising Hamiltonians. Whether these features are more likely in instances of a particular model, is a different question. Nevertheless, we will shortly summarize some known results on the Sherrington-Kirkpatrick model in the following.

We start with the observation, that the model in Eq.~\eqref{eq: SK model} is not extensive. The energy density will grow with system size due to the mean-field nature of the interaction. To make it extensive the bonds have to be scaled with $1/\sqrt{N}$. We use the non-extensive version since this seems to correspond to the situation in quantum annealers and therefore such models are used in the relevant literature~\cite{katzgraber_seeking_2015}. This however means, that one should be careful when comparing results for different system sizes, although the variation in our sizes is not extremely large when taking the square root. In the extensive version, the model with Gaussian bonds has a thermal spin glass transition at $\beta_{\mathrm{crit}}=1$~\cite{sherrington_solvable_1975, thouless_solution_1977, fischer_spin_1991}, which is expected to be modified only slightly, if at all, for bimodal bonds\footnote{Both distributions have the same variance, which is the relevant quantity for many results on the transition. However, other properties of course differ.}. In our version this has to be scaled by $1/\sqrt{N}$. The ground state energy density of the bimodal normalized model has been investigated in~\cite{boettcher_extremal_2005}. Converted to our version it scales as 
\begin{equation*}
\frac{E_{0}}{N}\approx 0.76\sqrt{N}+aN^{-1/6}.
\end{equation*}
Finally, in a transverse field the normalized model with Gaussian bonds has a quantum phase transition at $s_{\mathrm{crit}}\approx 0.397$~\cite{ andreanov_longrange_2012}. To use this result for our version, we look at how the versions transform between each other
\begin{align*}
(1-\tilde{s})\opnamed{H}{x}+\frac{\tilde{s}}{\sqrt{N}}\opnamed{H}{SK} &= A(\tilde{s})(1-s(\tilde{s}))\opnamed{H}{x}+s(\tilde{s})\opnamed{H}{SK}\\
\Rightarrow  A(\tilde{s}) &=1+\tilde{s}\left(\frac{1}{\sqrt{N}} - 1\right)\\
\Rightarrow  s(\tilde{s}) &=\frac{\tilde{s}}{A\sqrt{N}}=\frac{\tilde{s}}{\sqrt{N}+\tilde{s}\left(1 - \sqrt{N}\right)}.
\end{align*}
Here, we see that in our version the value of $s$ corresponding to a given $\tilde{s}$ in the normalized version decreases with $1/\sqrt{N}$. Therefore, the quantum phase transition will be pushed to $s_{\mathrm{crit}}=0$ in the thermodynamic limit, while lying around $s_{\mathrm{crit}} \approx 0.19-0.12$ for sizes $N=8-22$. Through the transformation, static properties of the versions can be related, thus the properties of the instances, such as the minimal gaps, will also occur with normalization. For the dynamic properties, the situation is a bit more difficult since through the non-linear nature of the transformation, the schedule of $s$ i.e. $s(t)=t/T$ will not transform into a similarly simple 
schedule for $\tilde{s}$. Therefore, we make no claims concerning the dynamics of the normalized model.

In the Ising limit ($s \to 1$) the eigenstates are product states in the $z$-basis. The energy differences are multiples of $2$, due to the discrete values of the couplings ($J_{ij}=\pm 1$). As a result, the energy difference from any bond, when changing one of the connected spins is $\pm 2$. Since the spectral width grows as $\sqrt{N}$ as discussed earlier, while the Hilbert space dimension scales as $2^{N}$, one can expect a large degeneracy for each energy. As we will discuss in the next subsection though, we focus on the case of a ground state which only features the always present double degeneracy due to spin-flip symmetry. Nevertheless, we typically expect the first excited state to be largely, perhaps even exponentially degenerate, for some instances. In the literature~\cite{altshuler_anderson_2010, knysh_relevance_2010}, perturbation theory in the Ising limit for similar problems with a large degeneracy in the excited sector was discussed with inconclusive results, concerning the range of applicability and the predictions for minimal gaps, to our best understanding.

Generally, a large difficulty is that the transverse field term, which acts as the perturbation, only has matrix elements between states with a single flipped spin. Another characterization of this situation is, that the states have a \emph{Hamming distance} of $1$, where the Hamming distance is the number of flipped spins. However, it is very well possible, that many states in the first excited sector have a large Hamming distance to the ground state and between themselves. In fact, there are many indications that a large Hamming distance between the ground state and the first excited state(s) is common for hard instances~\cite{boixo_evidence_2014, mehta_hardness_2022}. The order of the perturbation theory is then accordingly large and the convergence difficult to analyze. A more recent work~\cite{knysh_zerotemperature_2016}, used a different approach for the \emph{Hopfield model} (a different spin glass model) and found that on the Ising side of the phase diagram, there are multiple small gaps, where the gap size scales with system size as a stretched exponential, before scaling featuring a power law scaling at the transition point, consistent with a second order transition. For the hard instances in our example, to be discussed in the next subsection, we also see the minimal gap at values of $s$ deviating from the approximated transition point from the earlier discussion. However, ultimately we can not tell, whether this is due to the same reasoning as in the reference or whether the finite-size effects and/or corrections from the bimodal distribution, as opposed to the Gaussian distribution, of the bonds.
\subsection{\label{ssec: annealing model - instances}Instances}
\begin{figure*}
\centering
\includegraphics[width=0.9\linewidth]{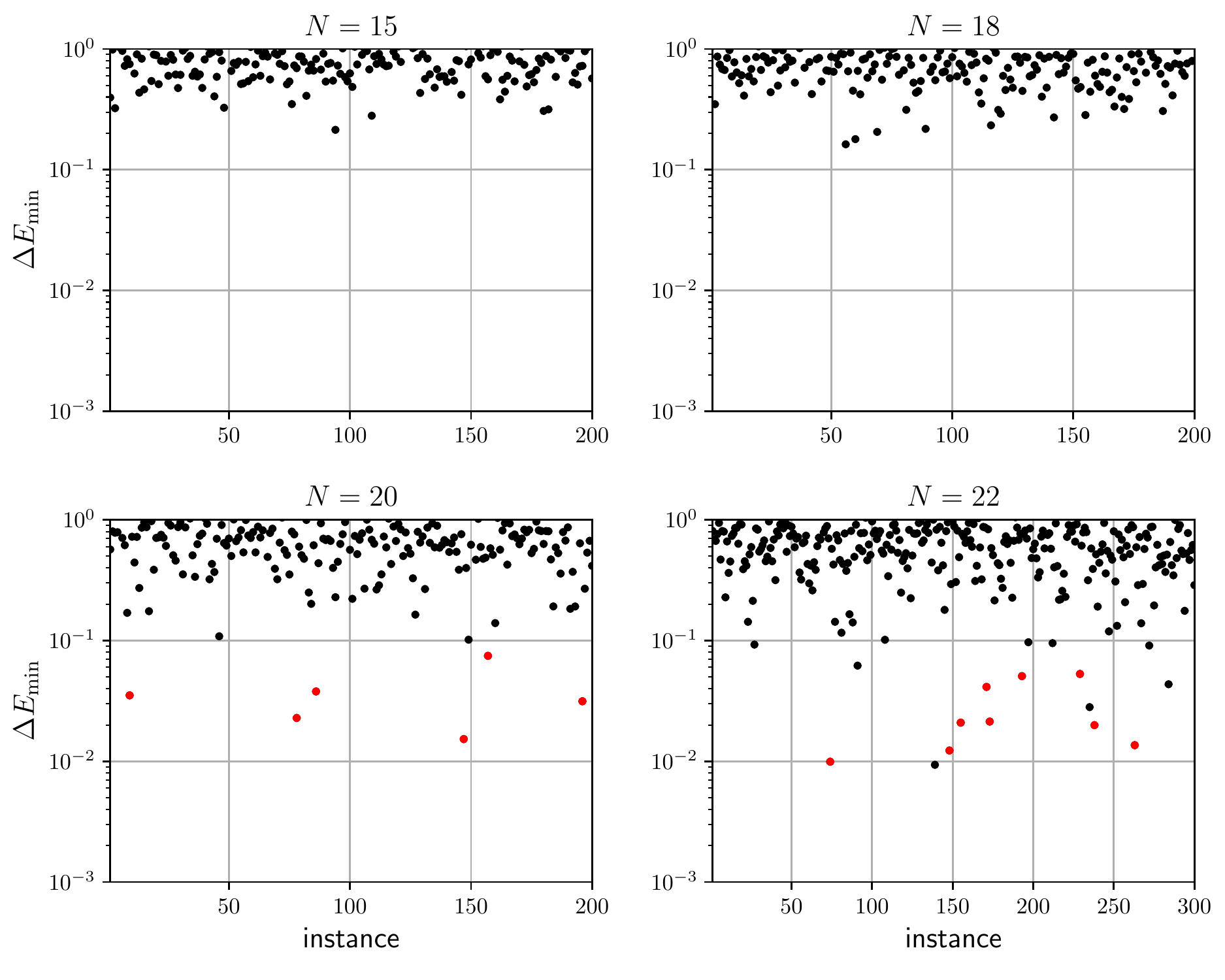} 
\caption{\label{fig: annealing minimal gaps}Minimal gaps between the ground state and excited state of all instances for system sizes $N=15, 18, 20$ and $22$. The gaps are based on a resolution of $\Delta s = 0.01$. The colored circles correspond to the instances identified as hard as described in the text. As we can see, those also correspond to instances with the smallest minimal gaps, with some additional instances at $N=22$ having small gaps.}
\end{figure*}

\begin{figure*}
\centering
\includegraphics[width=0.9\linewidth]{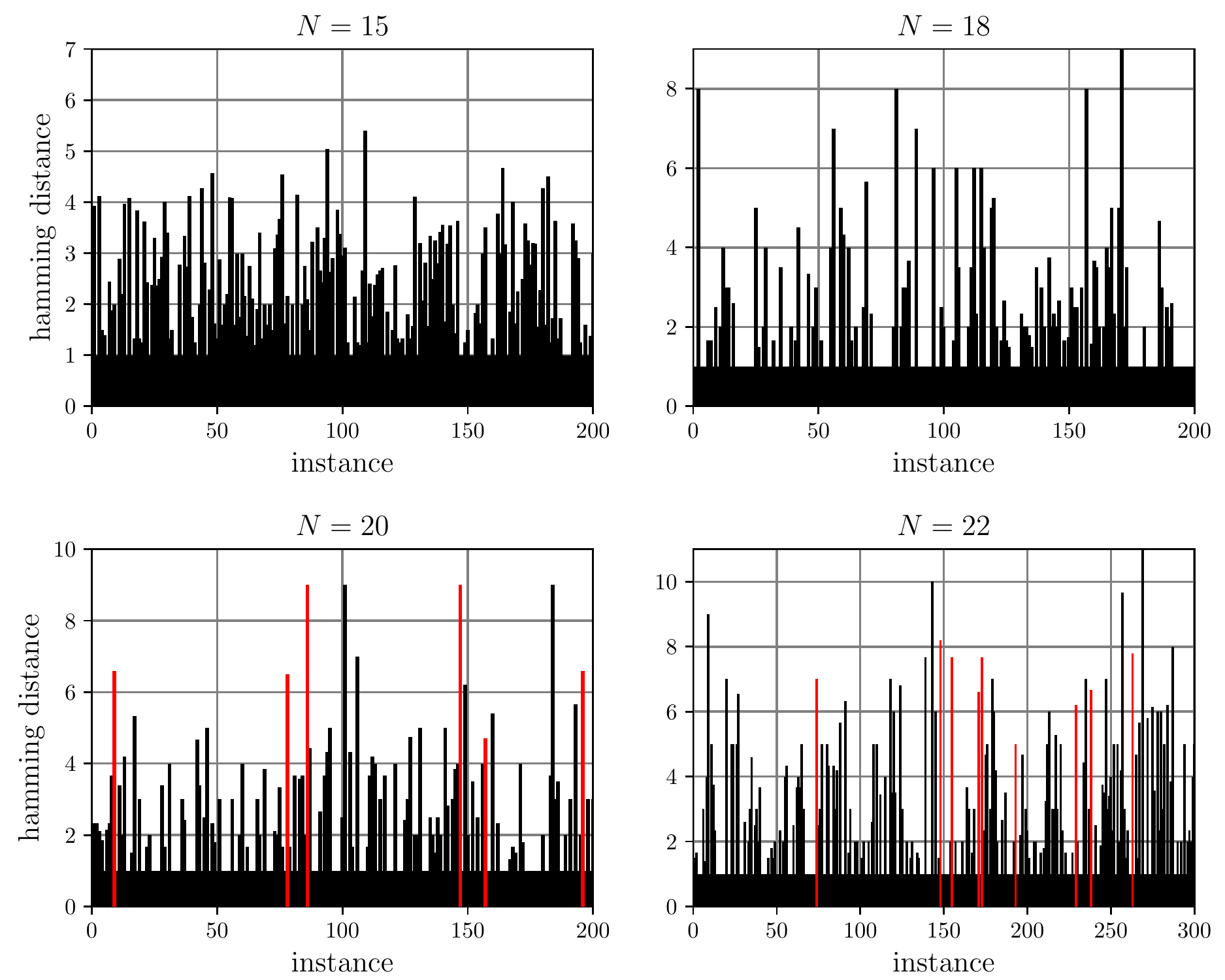} 
\caption{\label{fig: annealing hamming distance}Average Hamming distance between the ground state and states in the first excited sector of all instances for system sizes $N=15, 18, 20$ and $22$. The red bars correspond to the identified hard instances. The distance lies between $1$ and $N/2$, and not $N$, since the states have a spin-flip symmetry, and we take the minimal distance of each pair. As is evident from the colored bars, all identified instances have relatively large average distances, but there also exist other instances with a large distance. However, most instances have a smaller distance. This picture remains similar if the second excited sector is included.}
\end{figure*}

\begin{figure*}
\centering
\includegraphics[width=0.98\textwidth]{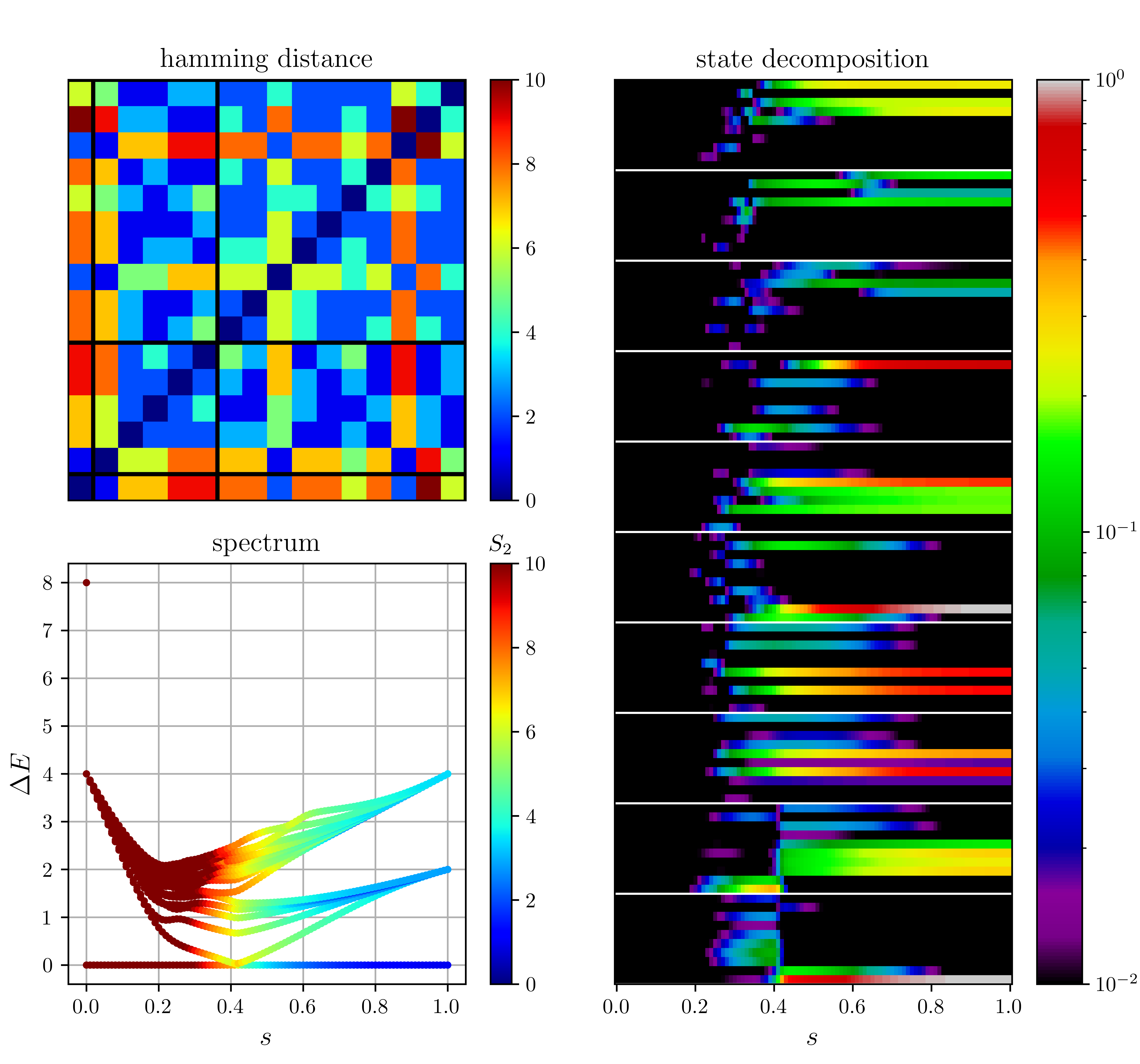} 
\caption{\label{fig: annealing spectrum N20 inst 9}Spectrum of a hard instance for $N=20$. The upper left plot shows the Hamming distance between the eigenstates of the Ising Hamiltonian within the first two degenerate subspaces. The lines indicate the subspace boundary; the lower left corner is the ground state. The lower left plot shows the spectrum, relative to the ground state, of a number of states corresponding to the subspaces. We see a very small gap between the ground and the first excited state around $s=0.4$. The color shows the entropy, in base $2$, of the eigenstates in the Ising basis. At the minimal gap, a fair number of states contributes to the ground and first excited states, which could imply limited applicability of perturbation theory. The right plot, showing the decomposition of the ten lowest states between themselves reveals though, that at least the ground state has a small contribution at the Ising side of the minimal gap.}
\end{figure*}

Having discussed the underlying model briefly, we now focus on an overview of the concrete instances, which were investigated. We have studied instances for various system sizes up to $N=22$ using quantum annealing, simulated annealing, and simulated quantum annealing. We focused mostly on even system sizes with an outlier in $N=15$ acting as a "bridge" between the smallest and largest sizes. In principle, it does not matter if the system size is even or odd, however, for odd sizes there can be "free spins" i.e. spins that in the ground state have the same number of satisfied and unsatisfied bonds. These free spins cause a degeneracy, therefore the fraction of configurations with degenerate ground states may be larger for odd sizes~\cite{rieger_quantum_2005}. The instances were chosen at random, however, only instances with a unique ground state (modulo spin-flip) were selected. Furthermore, we checked that the instances do not have simply permuted bonds, by comparing the full spectrum of each instance. It is unclear (to us), whether the restriction to a non-degenerate ground state (modulo the spin-flip symmetry) changes the complexity, but~\cite{katzgraber_seeking_2015} suggests it does not. Among these, we also identified "hard" instances based on the appearance of a \emph{diabatic bump} in intermediate-time quantum annealing, to be discussed in detail in Sec.~\ref{ssec: annealing fidelity - quantum annealing}. The number of instances and hard instances is summarized in table~\ref{tab: instances}.
\begin{table}
\centering
\begin{tabular}{c|c|c}
N & instances & "hard" \\ 
\hline 
8 & 100 & - \\ 
\hline 
10 & 100 & - \\ 
\hline 
12 & 200 & - \\ 
\hline 
15 & 200 & - \\ 
\hline 
18 & 200 & - \\ 
\hline 
20 & 200 & 6 \\ 
\hline 
22 & 300 & 9 \\ 
\end{tabular} 
\caption{\label{tab: instances}Number of investigated instances and identified hard instances for each system size.}
\end{table}
Other measures of hardness include solution time and the minimal energy gap~\cite{albash_adiabatic_2018}. The hardness also was found to correlate with a large Hamming distance between the ground and first excited states of the model~\cite{boixo_evidence_2014}. These measures seem to be mostly consistent with our definition, for example, the minimal energy gaps for each instance of the larger system sizes plotted in Fig.~\ref{fig: annealing minimal gaps} show, that the identified instances are among those with the smallest gaps. Also, for the smaller sizes, for which we do not find hard instances, the variation in the gap sizes seems much smaller, so the absence of hard instances is supported by the data.
Concerning the Hamming distance, a similar picture emerges. The average distances between the ground state and the first excited sector for each instance are shown in Fig.~\ref{fig: annealing hamming distance}. Note, that while the Hamming distance ranges between $0$ and $N$ in principle, due to the spin-flip symmetry the fully flipped states with distance $N$ are equivalent. Therefore, we use the minimal distance between the \emph{pairs} of states, which lies between $0$ and $N/2$. Of course, it is only $0$ for the ground state itself, which is not included in the averaging. The main conclusion is the same as for the gaps. None of the identified hard instances have a small average distance, while other instances with a large average distance exist. Most instances have a small average distance though. This remains true if we consider the distance of the ground state to the first excited sector or to the first and second sector combined, which we do not show here explicitly.
We would like to end the discussion of the model with a figure, visualizing some of the aforementioned concepts for a single hard instance with $N=20$ in Fig.~\ref{fig: annealing spectrum N20 inst 9}. The figure consists of three main parts based on the states in the first two excited subspaces of the instance i.e. states including the ground state, all first excited states, and all second excited states of the Ising part.

In the upper left plot, the Hamming distance between the eigenstates of the Ising part is plotted as a matrix: the lower left corner corresponds to the ground state followed by the first excited sector and the second sector, with the states not having any particular order within the sectors. There are black lines indicating the boundary of the ground state and the first and second excited state sectors. The coloring, corresponding to the distance, reveals that for this particular instance many states in the first and second sector have a large distance to the ground state, while the distances within and between the excited sectors seem to vary, but mostly to be on the smaller side. We find it noteworthy, that there is a state with a very small distance to the ground state in the first sector, which also features rather large distances to other states in the sector. We cannot tell, whether this scenario leads to hard instances in general, but at least it illustrates the complexity of a perturbation theory and strengthens the intuition that the Hamming distance is relevant, as we will see shortly. 

The lower plot on the left shows the (instantaneous) spectrum of a number of states corresponding to the first sectors as before~\footnote{We formulate carefully, since this does not need to be the same number as in the first sectors in the transverse field limit.}. These were obtained using a sparse matrix algorithm from \emph{SciPy}, which computes a specified number of eigenstates and eigenenergies. The energies are measured with respect to the instantaneous ground state energy, which therefore always lies at $0$. The energies in both limits are discretized with the lowest excitation energy being $4$ in the transverse field limit and $2$ in the Ising limit. As expected from a hard instance, the minimal gap at $s \approx 0.4$ is rather small, in fact not visible by the eye. This is different from many other instances. The color here is based on the entropy in base $2$ computed from the weights of the eigenstates in the Ising basis. Naturally, the entropy is high in the transverse field limit, since therein the eigenstates are almost equal weight superpositions. In the Ising limit, the entropy is lower, but non-vanishing, indicating that the eigenstates at a low transverse field are superpositions of a fair number of Ising eigenstates. Close to the minimal gap, the entropy of individual states changes relatively rapidly with an intermediate value in the vicinity of the minimum. This suggests that a significant number of Ising states is involved and as a consequence, the usefulness of perturbation theory may be limited here. 

Finally, in the right part of the figure, we decompose each of the $10$ lowest instantaneous eigenstates in the basis of the $10$ lowest Ising eigenstates, with the same order as in the Hamming distance plot. The (logarithmic) color scale shows these weights as a function of $s$. The plot is divided into segments separated by white lines, with each segment corresponding to an eigenstate with the lowest (the ground state) being at the bottom and energy increasing to the top. Within each segment, the weights are in the same order. For example, the lowest colored bar shows, that the instantaneous ground state has a high overlap with the Ising ground state on the right side of the minimum, also mixing slightly with the lowest distance excited state mentioned earlier. On the left side, it is a mixture of several states, not including the lowest distance state. The mixture fits well to the states contributing to the first instantaneous excited state, as seen in the second segment from the bottom. In case this is not clear, the state in the second segment also corresponds to the state with the lowest energy above the ground state in the spectral plot. Here one can see clearly, that the first excited state on the Ising side consists of states with a large Hamming distance to the ground state explaining the very small energy gap at the minimum.
With this, we finish the, admittedly exhaustive, discussion of Fig.~\ref{fig: annealing spectrum N20 inst 9} and the static properties of the model. In particular, the interplay between the instantaneous and the Ising basis can be useful in understanding dynamical properties, which we turn to in the next section.
\section{\label{sec: annealing fidelity}Fidelity}
In the next sections, we will discuss different quantities from the simulations. We start with the most relevant quantity for optimization, which is the fidelity $\mathcal{F}$ i.e. the probability of finding the ground state. We will start with a thorough discussion of quantum annealing and then comment on both classical algorithms.
\subsection{\label{ssec: annealing fidelity - quantum annealing}Quantum annealing}
\begin{figure}
\subfloat{\includegraphics[width=0.98\linewidth]{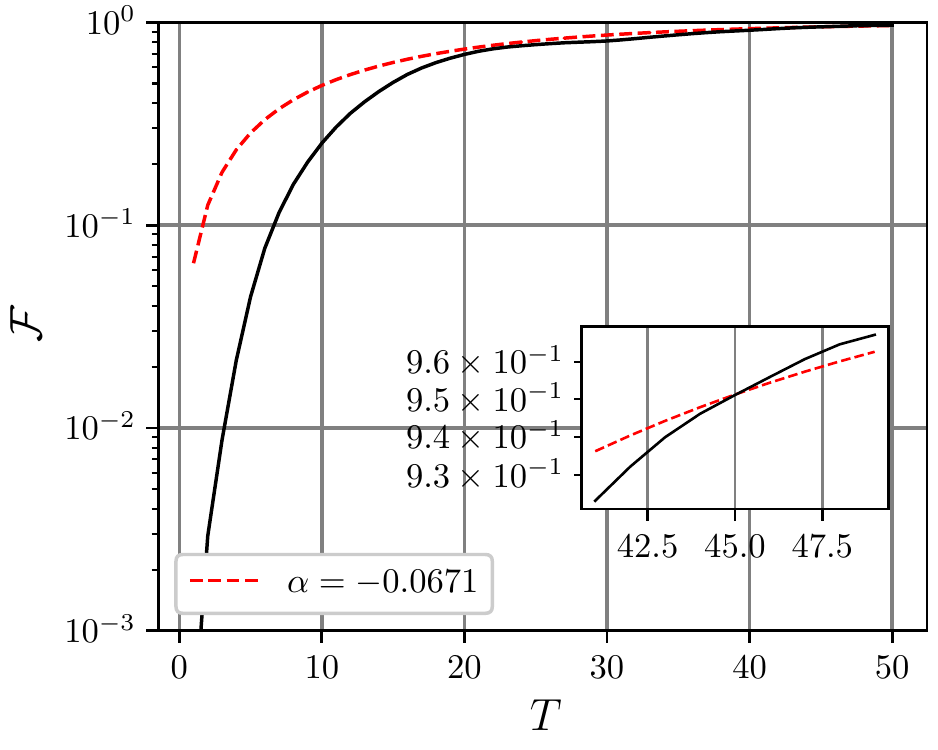}}\\
 \subfloat{\includegraphics[width=0.98\linewidth]{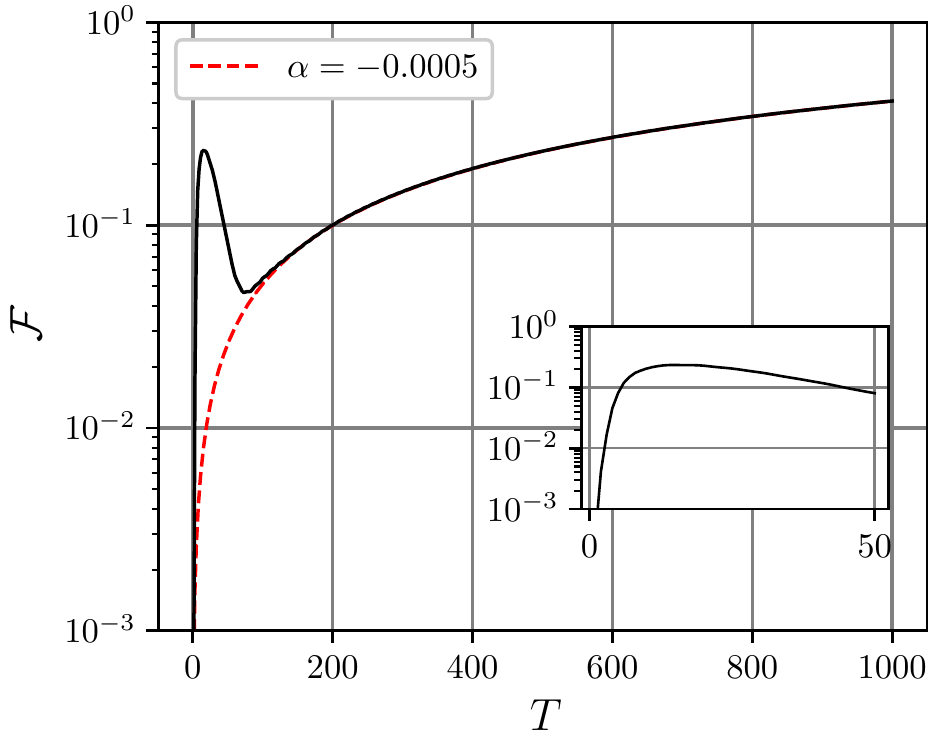}}
\caption{\label{fig: qa fidelity single instance}Fidelity as a function of the annealing time for $N=20$. Shown are a normal instance (upper) and a hard instance (lower). Exponential fits to larger annealing times are plotted in red, with the exponent $\alpha$ indicated in the legend. The normal instance reaches an almost unit fidelity at $T=50$ with the fidelity being only roughly approximated by an exponential. For the hard instance unit fidelity is not yet reached even for $T=1000$, however, seems to be well approximated by an exponential asymptotically. The fidelity of the normal instance grows monotonically, while the hard instance features a diabatic bump at intermediate times. The inset, showing the same time window as for the normal instance, suggests that before the bump the fidelity is around $10 \%$ in both instances.}
\end{figure}

Let us begin the discussion, by analyzing the dynamics of a hard and a "normal/simple" instance; the hard instance is the same as in Fig.~\ref{fig: annealing spectrum N20 inst 9}. In Fig.~\ref{fig: qa fidelity single instance}, the fidelity at the end of an annealing run is plotted as a function of the annealing time $T$ for $N=20$.
The upper plot shows the normal instance and the lower the hard one.\\
Additionally, the fidelity at larger annealing times is fitted to an exponential form $\mathcal{F} \sim 1 -\erm^{-\alpha T}$ and the fit exponent in each case is shown in the legend. For the normal instance, we observe an (almost) monotonic increase of the fidelity with time, with unit fidelity almost reached in the time window up to $T=50$. 

The inset shows, that an exponential approximates the behavior only roughly at longer times. Since the plots are logarithmic, an exponential would show as a (strictly) linear behavior. For the hard instance, the situation is richer: at short times up to $T\approx 10$ the fidelity behaves comparably to the normal case, rising monotonically to around $10 \%$. This is well seen in the inset showing the same time window. Then, however, a striking difference appears signified by a very rapid decrease of the fidelity by about a factor of $5$. This fidelity peak, which we take as the defining feature of hard instances in this study, was observed in some earlier works~\cite{ crosson_different_2014, zhou_quantum_2020} and referred to as a diabatic bump. We will comment on the mechanism behind it and why we expect it to appear for hard instances a bit later. After the bump, we observe an exponential behavior persisting up to $T=1000$, at which point in time unit fidelity is still not reached for this instance.

\begin{figure}
\subfloat{\includegraphics[width=0.98\linewidth]{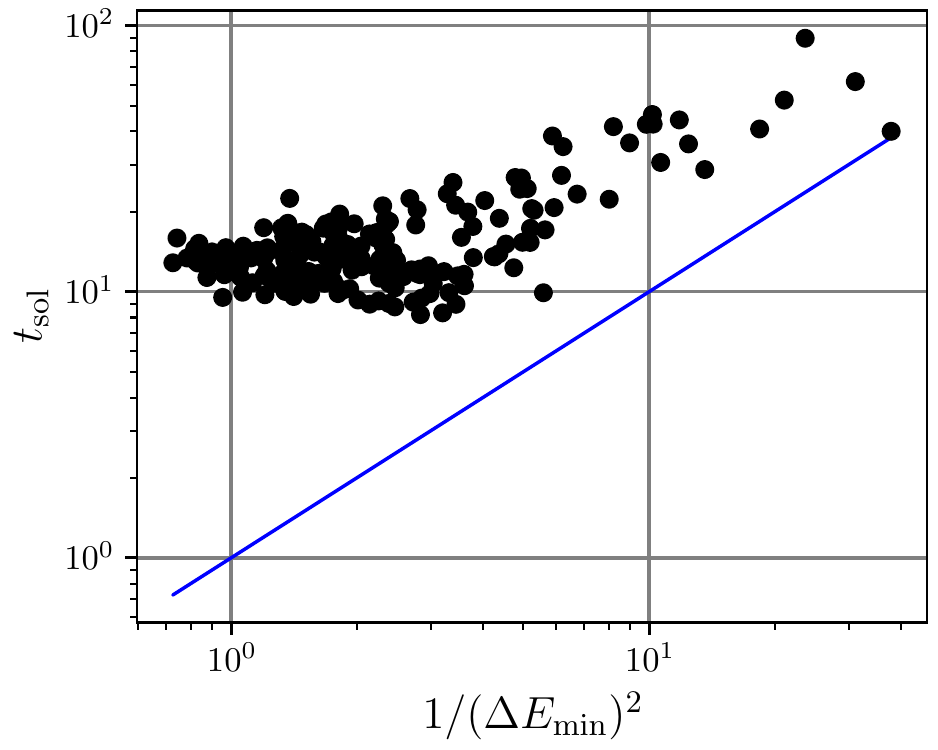}}\\
\subfloat{\includegraphics[width=0.98\linewidth]{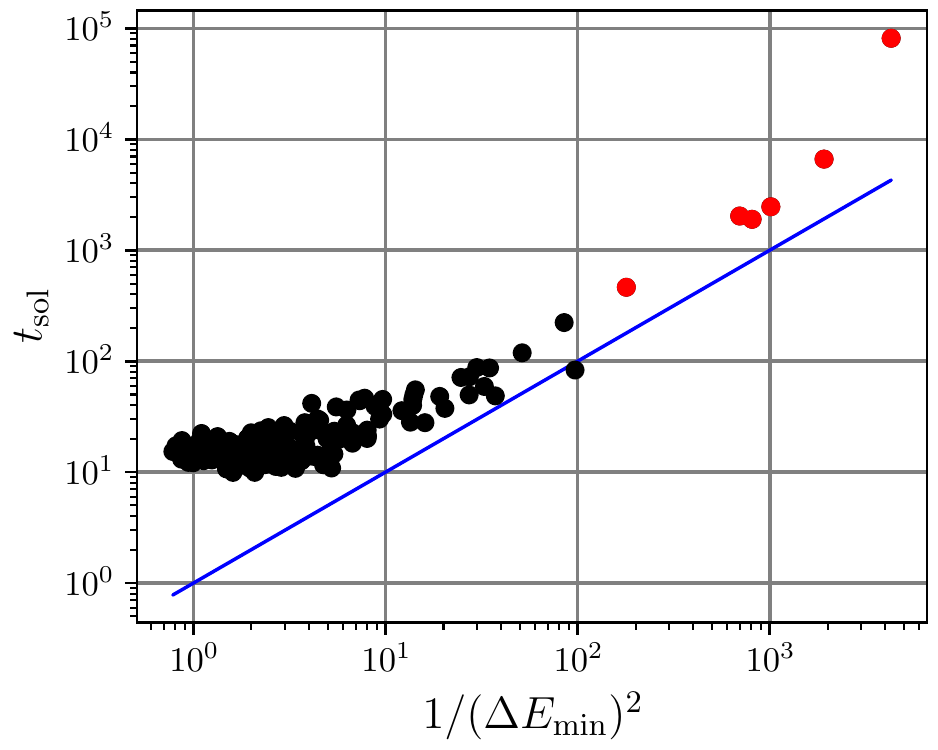}}
\caption{\label{fig: qa soltime vs gap}Solution times obtained from exponential fits to the fidelity dynamics, plotted against the inverse square of the minimal gap for each instance of system sizes $N=18$ (upper) and $N=20$ (lower). The colored circles represent the hard instances. The line in the background serves as a guide for the expected linear scaling. This scaling seems consistent with the data at small gaps, while at large gaps the solution times show no clear scaling, while still being small at those gap values.}
\end{figure}

We will explain why one would expect asymptotic exponential behavior asymptotically, at least for hard instances, later. Before, we note that if it manifests, one can define the time to solution $\opnamed{t}{sol}$ sensibly, by the inverse of the exponent $\opnamed{t}{sol}=1/\alpha$. As we again will see later, a scaling with the inverse minimal gap squared is expected for the solution time. We test this for two larger system sizes $N=18$ and $N=20$ in Fig.~\ref{fig: qa soltime vs gap}. Unfortunately, for $N=22$ we did not reach simulation times large enough to resolve the exponents well for hard instances. Here, we see the solution time as defined above, plotted against the minimal gap already discussed in Sec.~\ref{ssec: annealing model - instances} and shown in Fig.~\ref{fig: annealing minimal gaps}. For the largest gaps, the solution times are also small, but there is not any visible scaling. This is not surprising, since we have already seen that the exponential is only a very rough fit for instances with small solution times. But in any case, small solution times are still visibly related to large gaps. For smaller gaps, a trend consistent with an inverse square scaling is discernible visually though, guided by the exact line in the background. Especially the hard instances colored in red, follow the trend rather well. Here, one should also keep in mind that the smallest gaps may also be limited by the resolution as discussed in Sec.~\ref{ssec: annealing model - instances}.

In Fig.~\ref{fig: qa fidelity single instance} and the corresponding discussion, we have seen three different behaviors of the fidelity: rapid increase at short times, diabatic bump at intermediate times for hard instances, and asymptotic exponential growth at long times, at least for hard instances. In the following, we will explain, how to interpret the intermediate and long-time behavior, while a discussion of the short-time behavior will be delayed until Sec.~\ref{sec: annealing dist}, where we will analyze the entire distribution using short-time expansions.
\subsubsection{\label{sssec: annealing fidelity - qa dia bump}Intermediate times - diabatic bump}
\begin{figure*}
\centering
\includegraphics[width=0.98\linewidth]{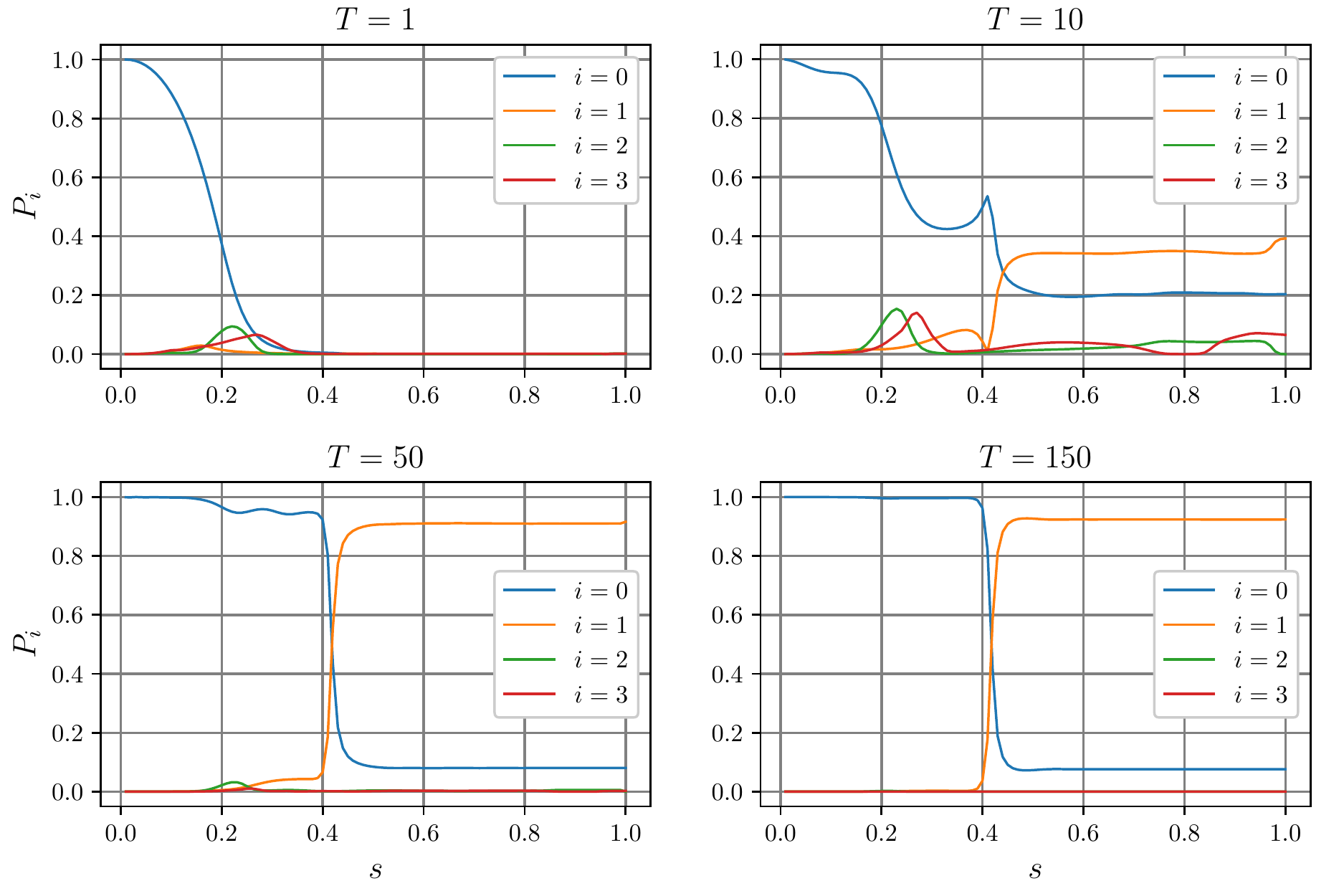} 
\caption{\label{fig: qa diab bump overview}Probabilities of the four instantaneous eigenstates with the lowest energy during annealing runs with varying annealing time $T$, for a hard instance of size $N=20$. }
\end{figure*}

As we have seen, some instances, which we refer to as hard instances, feature a peak in the fidelity as a function of the annealing time, in contrast to the (almost) monotonic growth in other instances. How can one interpret this phenomenon? As a starting point, we will take the same instance, now analyzing the full time-evolution during a protocol with annealing time $T$. In Fig.~\ref{fig: qa diab bump overview}, the time-evolution of the weights with respect to the four instantaneous states with the lowest energies are shown. These are obtained, by decomposing the full time-evolved state in the instantaneous eigenbasis using a Krylov method with an order of $40$. 

In the figure, four different annealing times are shown, starting at a short time $T=1$, moving to a time around the beginning of the bump $T=10$, transitioning to a time around the end of the bump $T=50$ and finishing with a time well after the bump $T=150$. For all annealing times, the ground state probability starts at $1$ and all other states have a probability of $0$. This is due to the preparation of the initial state, which is the ground state of the transverse field at $s=0$. The dynamics then leads to transitions between the states, depleting the instantaneous ground state over time. For the shortest time, it is fully depleted towards the end of the protocol along with the other low excited states, indicating that the probability distribution is spread over many states, presumably including higher excited states as well. For $T=10$, the depletion at the beginning is slower, but around $s=0.4$ there is a sudden drop in the ground state probability, accompanied by a rise in the probability of the first excited state. The occurrence is related to the minimal gap between being located at this value of $s$, as seen previously in Fig.~\ref{fig: annealing spectrum N20 inst 9}. Note that the "spiky" features of the probabilities just at the minimal gap may be remnants of the numerics, stemming from the need to resolve two states lying very close in energy at that point, but we expect the curves outside of the immediate vicinity to be accurate. For $T=50$ and $T=150$, the same observations can be made, where the ground state gets depleted increasingly less before the minimal gap, and there is a rapid transition at the minimal gap point, whereafter the probabilities are almost constant. Two further observations will help to understand the bump: with increasing time higher excited states play a lesser role in the dynamics. Already at $T=50$, they are almost invisible in the plot. Furthermore, we notice that the ground state probability for $T=10$ at the end of the protocol is \emph{higher} than for $ T=50 $ while being \emph{lower} before the minimal gap. For the transition, the initial ground state and excited state occupations, as well as the annealing time, are relevant. The exact dependence is presumably very complicated since at small times multiple states are involved. The annealing time could also have two competing effects. On one hand, it sets the time to traverse the gap and therefore the time effectively available for a transition. On the other hand though, at long times the transition will be suppressed with increased annealing time and the system remains in the ground state, even though the available time for a transition is plenty. The rough explanation for the diabatic bump based on these observations seems to be the following: with increasing annealing time the ground state probability before the minimal gap increases, while the first excited state remains at a low probability. At the minimal gap, the transition amplitude between the ground and excited state is large, resulting in a rapid transition. Before the bump, the ground state probability is relatively small and the effective transition time possibly limited. Here, the macroscopic fidelity of the first excited state before the avoided crossing is also an important factor as noted in~\cite{crosson_different_2014}, leading to a macroscopic fidelity in the ground state after the crossing. As seen in the figure, this excited state fidelity increases with increasing time at small times. During the bump, the initial probability will increase as will the effective transition time, such that the final ground state probability is lower with increased annealing time. Only once the annealing time is large enough to cause a suppression of the transition, the final probability increases again with annealing time. We will discuss a possible mechanism for the suppression at large times as well as the exponential growth of the fidelity in the next subsection.
Before, we note that the process can also be viewed from another perspective, using a fixed basis instead of the instantaneous basis. Here, transitions between states are suppressed at short times, since the state then does not have enough time to change significantly. In fact, for infinitely short times, the state simply remains constant (modulo a possibly acquiring a phase). This situation is also known as a quench or the \emph{sudden approximation}. Since the initial state is not the ground state of the final Hamiltonian, the fidelity at the end is small in this case. At the minimal gap, a narrow avoided crossing occurs; the states in the constant basis are the same before and after the avoided crossing, but their order is changed. Hence, if the state being close to the ground state initially changes only slightly, it is close to the first excited state afterward. Only at long times the system has enough time to perform the transition between the ground state and excited state at the avoided crossing, resulting in the ground state after the avoided crossing.
\subsubsection{\label{sssec: annealing fidelity - qa landau-zener}Long times - Landau-Zener physics}
As we have seen, at longer annealing times, only the two lowest energy states in the instantaneous basis are involved in the dynamics. A two-level system with a changing basis can be treated analytically. In the literature, the model 
\begin{equation}
H(t)=\lambda t\pauli{z}{}+\Delta \pauli{x}{},
\end{equation}
is often referred to as the Landau-Zener model~\cite{vitanov_landauzener_1996, vitanov_transition_1999,  damski_simplest_2005, degrandi_adiabatic_2010a, sinitsyn_quest_2017}. The state is initially prepared in the ground state of the Hamilton at $t \to -\infty$ and the fidelity is measured with respect to the ground state at $t \to \infty$. Finite-time setups, more akin to our case, have been studied in~\cite{vitanov_landauzener_1996, vitanov_transition_1999}. We will not summarize the exact solution for all times, as it involves several special functions, but would like to note, that asymptotically the fidelity scales as 
\begin{equation}
\mathcal{F} \approx 1 - \erm^{-\pi \frac{\Delta^{2}}{\lambda}},
\end{equation}
where we note, that $\lambda$ being the "rate of parameter change" is roughly related to the inverse annealing time $1/T$. Since $\Delta$ is precisely the minimal gap in this model, we recover the inverse square dependence of the solution time.

Clearly, for a many-body system, the applicability is limited, since there are more possible transitions. There are studies of multiple levels, in the context of multi-state Landau-Zener models~\cite{sinitsyn_quest_2017}. To our best understanding though, the treatment, in that case, is significantly more involved, while still being limited to a relatively small number of levels. Yet, in the context of the adiabatic theorem, many results suggest a timescale dictated by the inverse square gap even for many-body systems~\cite{morita_mathematical_2008, degrandi_adiabatic_2010, bachmann_adiabatic_2017, albash_adiabatic_2018}, although these rely on bounds rather than exact solutions of the dynamics. Therefore, it is perhaps not surprising that the formula works well for the long-time dynamics of hard instances, where the minimal gap between the ground and excited state can be orders of magnitude smaller than gaps to the other levels. For simpler instances such a separation of scales is not given, therefore a multi-state description is probably necessary, even for long times.

We finish the subsection by reminding the reader, that the exponential behavior discussed in the section, appears for a set system size. The main question in the research on quantum and classical annealing is, how the corresponding timescale scales with system size. While for quantum annealing a rough classification is possible based on the order of the phase transition, the treatment of classical annealing is rather limited to our best knowledge. In this study, we do not investigate this question though, but rather focus on understanding the dynamics of selected hard or normal instances for fixed sizes across a range of annealing times.

\subsection{\label{ssec: annealing fidelity - classical annealing}Classical annealing}
\begin{figure}
\subfloat{\includegraphics[width=0.98\linewidth]{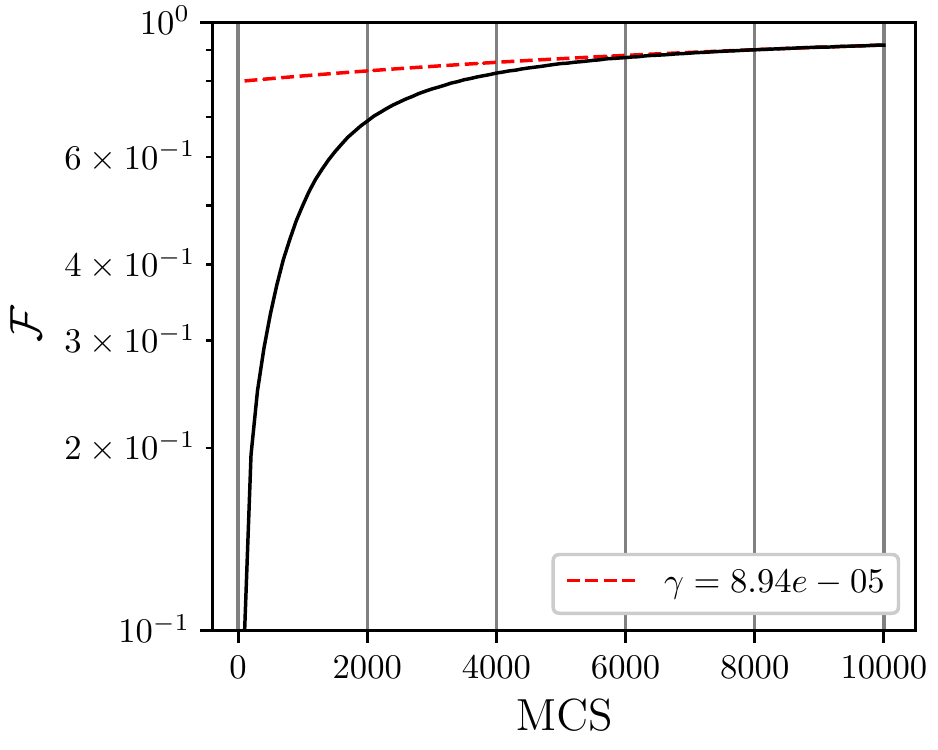}}\\
 \subfloat{\includegraphics[width=0.98\linewidth]{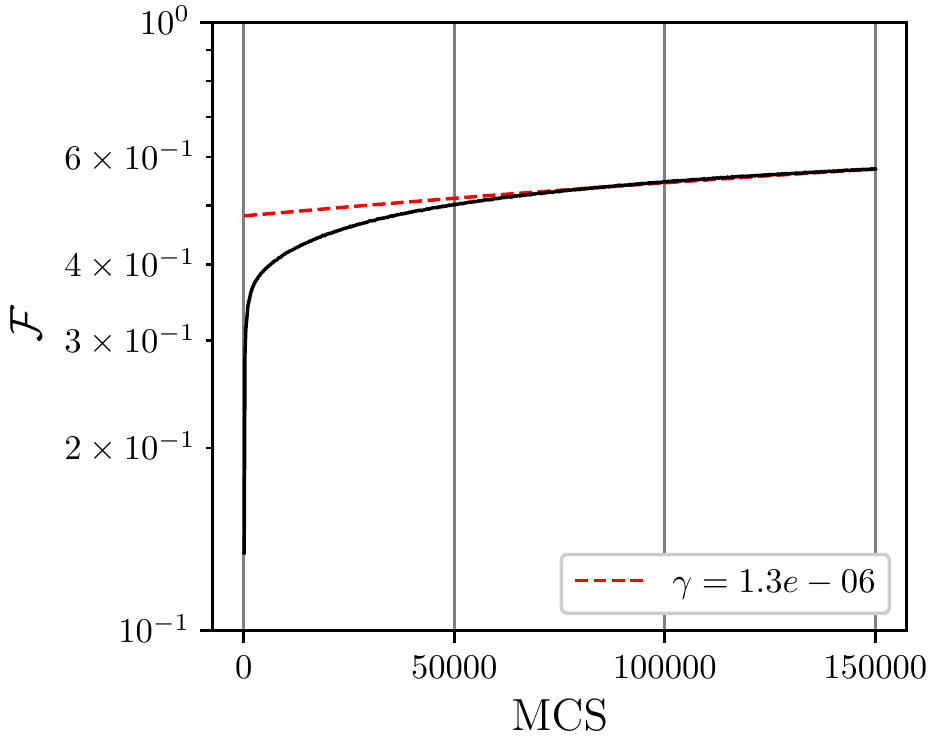}}
\caption{\label{fig: sa fidelity single instance}Fidelity as a function of the annealing time for $N=20$. Shown are a normal instance (upper) and a hard instance (lower). Exponential fits to larger annealing times are plotted in red, with the exponent $\gamma$ indicated in the legend. The normal instance lies around $90 \%$ after $10000$ Monte Carlo steps, while the hard one reaches around $40 \%$ quickly and features an extremely slow growth from thereon.}
\end{figure}

For both classical annealing methods, such qualitative differences between instances can not be observed. Instead, the fidelity grows monotonously, albeit at different rates depending on the instance. As an example, in Fig.~\ref{fig: sa fidelity single instance} the fidelity for the same instance as before is shown. 
As one can see, there is a monotonous growth with a roughly exponential behavior at long annealing times (measured in Monte Carlo steps). Theoretical approaches to estimate the timescale exist ~\cite{morita_mathematical_2008, nishimori_relation_2015, levin_markov_2017}, but we will not discuss them in detail here.
\section{\label{sec: annealing corrs}Correlation functions}
Important and experimentally accessible observables during the dynamics are the correlation functions
\begin{equation}
\label{eq: corrs zz}
G_{ij}=\braket{\pauli{z}{i}\pauli{z}{j}}.
\end{equation}
Since $\braket{\pauli{z}{i}}=0$ for all times due to symmetry, we do not distinguish between (un-)connected correlation functions. As a further motivation to study these objects, we note that Ising energy $\braket{\opnamed{H}{SK}}$ is fully determined by their values and that all spin-flipped pairs in the computational basis can be identified uniquely by their correlations. This can be done by choosing the sign of the first spin arbitrarily, then determining the sign of the second by the value of $G_{12}$ and continuing accordingly. The diagonal entries $G_{ii}=1$ and due to commutativity $G_{ij}=G_{ji}$, therefore we will only consider $i > j$ in the following. 

\begin{figure*}
\centering
\includegraphics[width=0.99\textwidth]{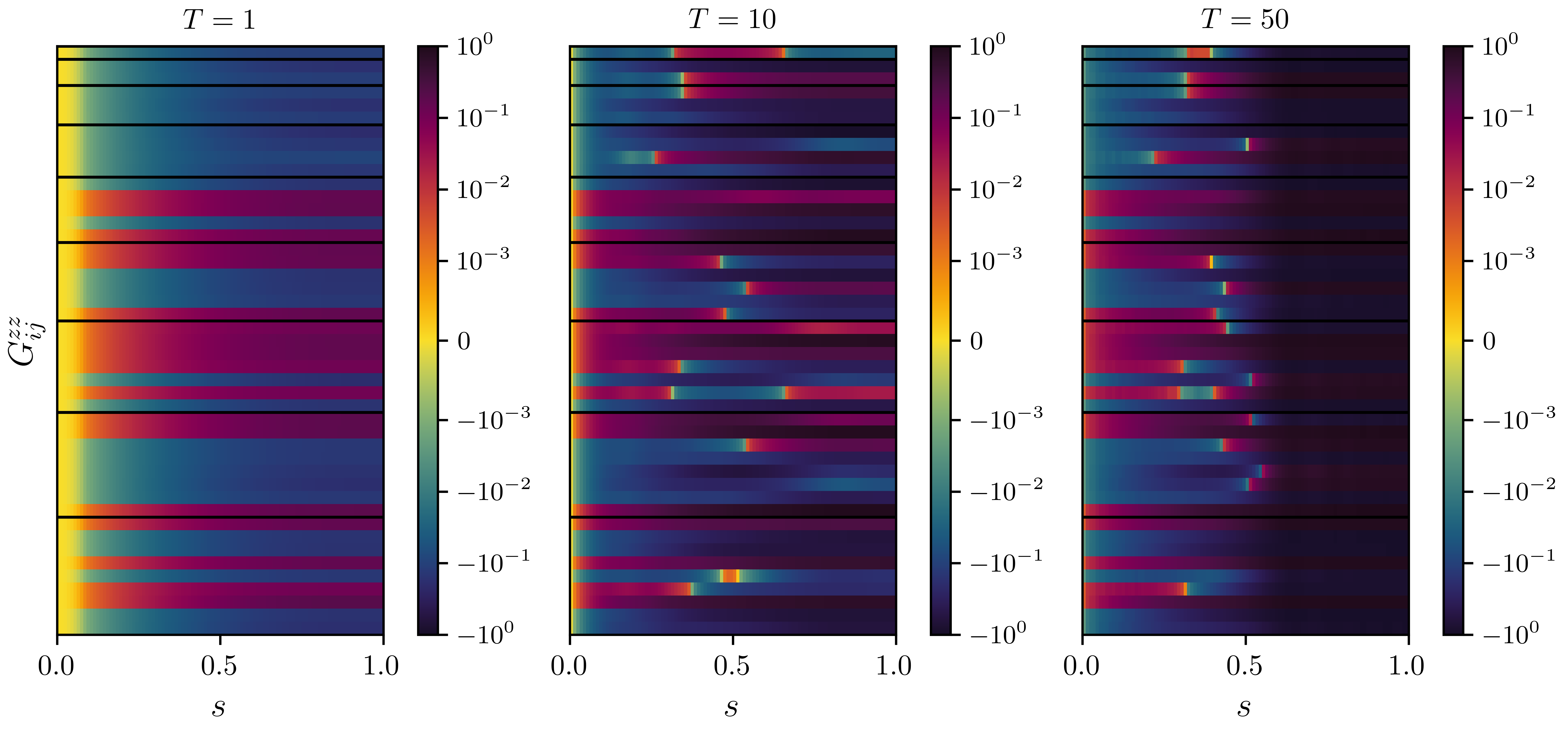}
\caption{\label{fig: correlations qa vary T}Correlations in quantum annealing during a protocol at different annealing times for an instance with $N=10$. Plotted are the correlations $G_{ij}$ for $i > j$, with the correlations of the first spin $G_{1\cdot}$ at the bottom and correlations of other spins in increasing order from bottom to top; the black lines indicate the different blocks $G_{i \bullet} \; $. For $T=1$ the correlations are determined by the bonds and their signs are $\sign{(G_{ij})}=-J_{ij}$. For $T=10$ we observe a sequence of sign changes. Finally, for $T=50$ there is a saturation phase, once the correlations reach their correct signs. Note that correlations with the incorrect as well as the correct initial signs can feature sign changes. Also, the location of the sign changes as well as the total number depends on the annealing time. See for instance, the double sign change at the bottom visible for $T=10$ is "closed" for $T=50$.}
\end{figure*}

In Fig.~\ref{fig: correlations qa vary T} the entire sweep for QA with varying annealing times is shown. Going through the sweep, we notice the following features, which seem to be general across our data:
\begin{enumerate}
    \item[$\blacktriangleright$] Due to the choice of the initial state, correlations vanish at the start of the protocol. For QA this is strict, while for the classical methods we expect a small finite value due to averaging over the finite number of runs. 
    \item[$\blacktriangleright $] For short times, correlations forming have the same sign structure for all methods. The signs are determined by the bonds, such that $\sign{(G_{ij})}=-J_{ij}$ . The term short-time here could mean a short sweep or a short portion of a long sweep. Later, in Sec.~\ref{sec: annealing dist}, we will show that these correlations follow from a short-time expansion and also appear in high-temperature expansions of the Ising Hamiltonian. 
    \item[$\blacktriangleright$] After some time, some bonds start to change their signs. Several sign changes can follow and the sign of a given correlation function can also change more than once. Correlations with the incorrect initial signs can change as well as those with correct initial signs.
    \item[$\blacktriangleright$] Once the signs are all correct, we do not observe further sign changes and during the remaining evolution, the magnitude saturates.    
\end{enumerate}

\begin{figure*}
\centering
\includegraphics[width=0.98\textwidth]{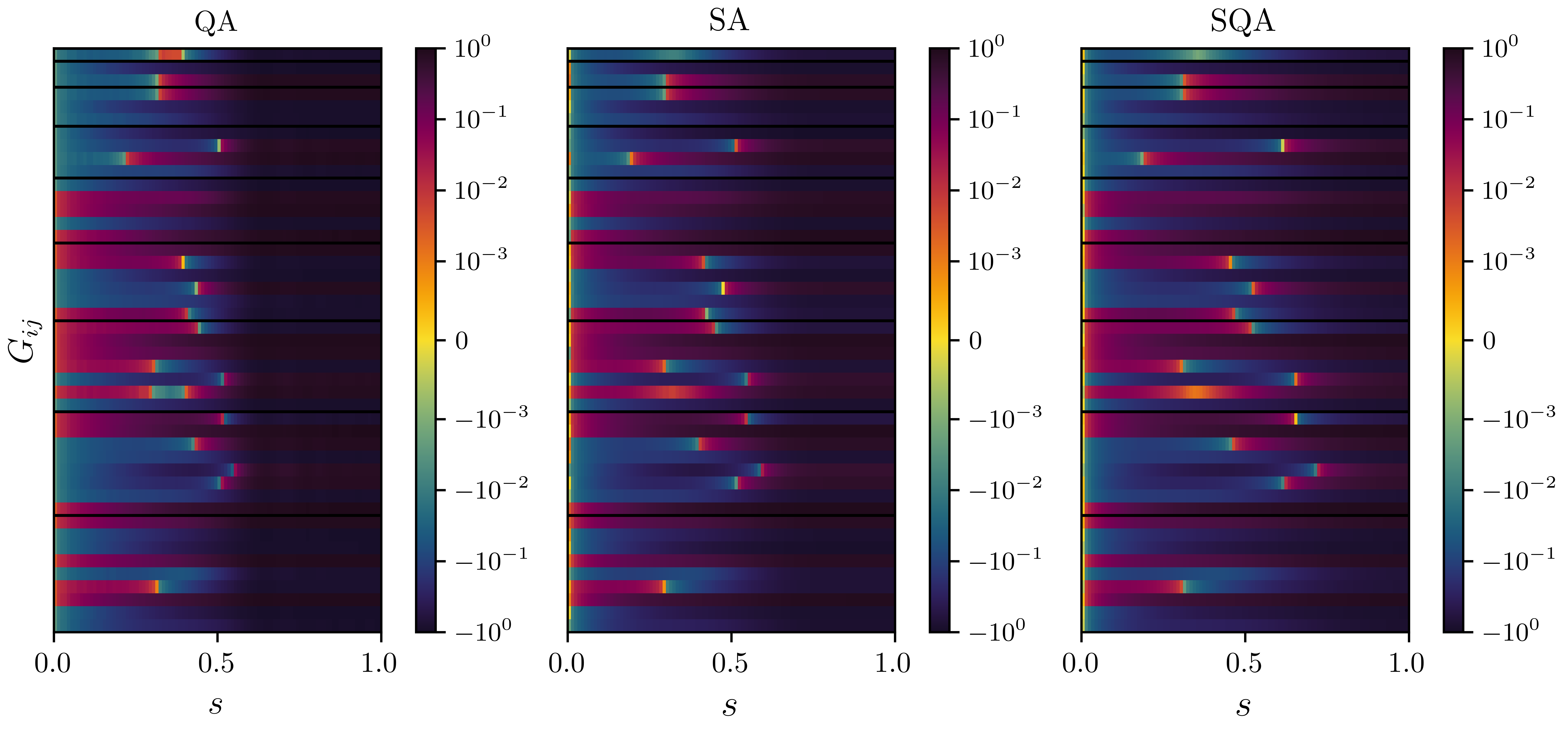} 
\caption{\label{fig: corr N10 compare}Comparison of quantum annealing, simulated annealing, and simulated quantum annealing of an instance with $N=10$ with annealing times of $T=50$ and $10000 \; \mathrm{MCS}$ in both classical methods. Plotted are the correlations during the protocol with annealing times sufficient to find the correct solution. The sign changes are remarkably similar for all three methods. The main differences are two double changes in quantum annealing, which however are observable in the other methods for shorter annealing times.}
\end{figure*}

This general behavior is also observed in simulated annealing and simulated quantum annealing. 
As an example, Fig.~\ref{fig: corr N10 compare} shows the sweep of the same instance as above at long times for the different methods.
A striking observation is, that the sign changes of the methods are very similar to each other. 
In fact the same signs change in roughly the same order, except for one double sign change in QA, which is actually present in the other methods at shorter annealing times. A sensible comparison may be possible between all three methods in the short-time limit, where however it is unclear if this limit extends to the first sign changes, which will be discussed in more detail in Sec.~\ref{sec: annealing dist}, as well as between QA and SQA in the very long-time limit. The latter can be justified by the fact that the correlations in QA are the ground state correlations at long times, which will also be the case for SQA, albeit formally only at zero temperature and an infinite number of replicas. The data suggest that at long times SA also has the same correlations, but this observation lacks a justification, as here the correlations in the long-time limit correspond to thermal correlations at the temperature set by $\beta(s)$. The reason is that at long times, the system should have enough time to equilibrate at every temperature along the sweep. 

\begin{figure*}
\centering
\subfloat{\includegraphics[width=0.9\linewidth]{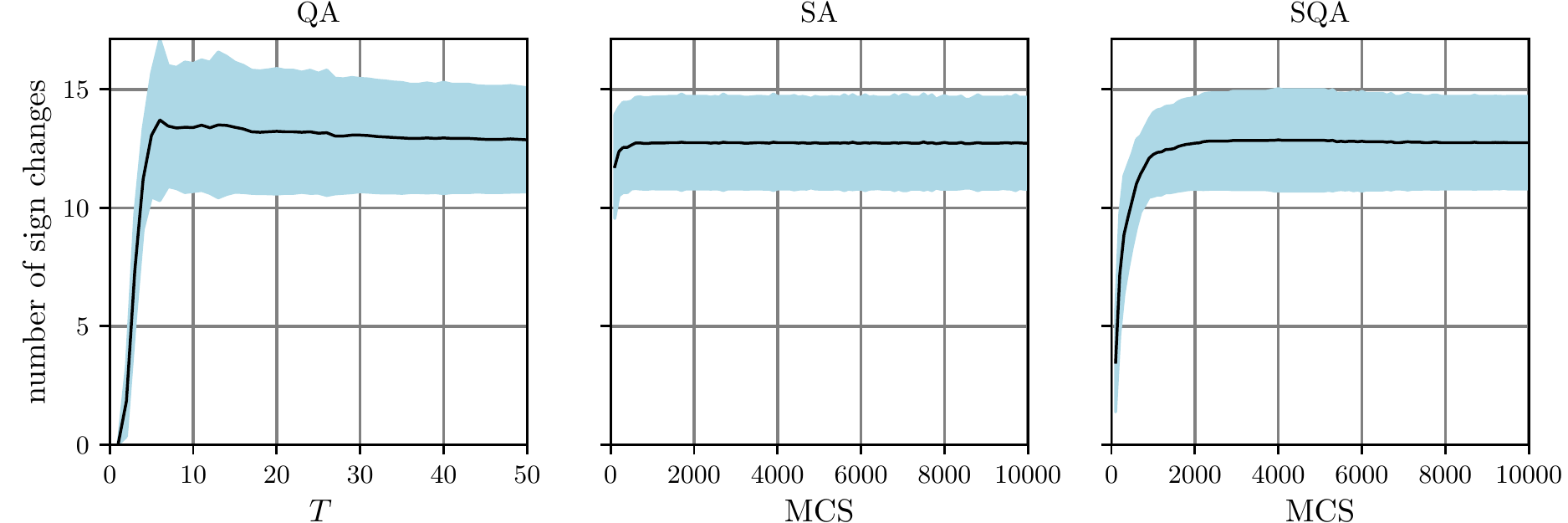}}\\
\subfloat{\includegraphics[width=0.9\linewidth]{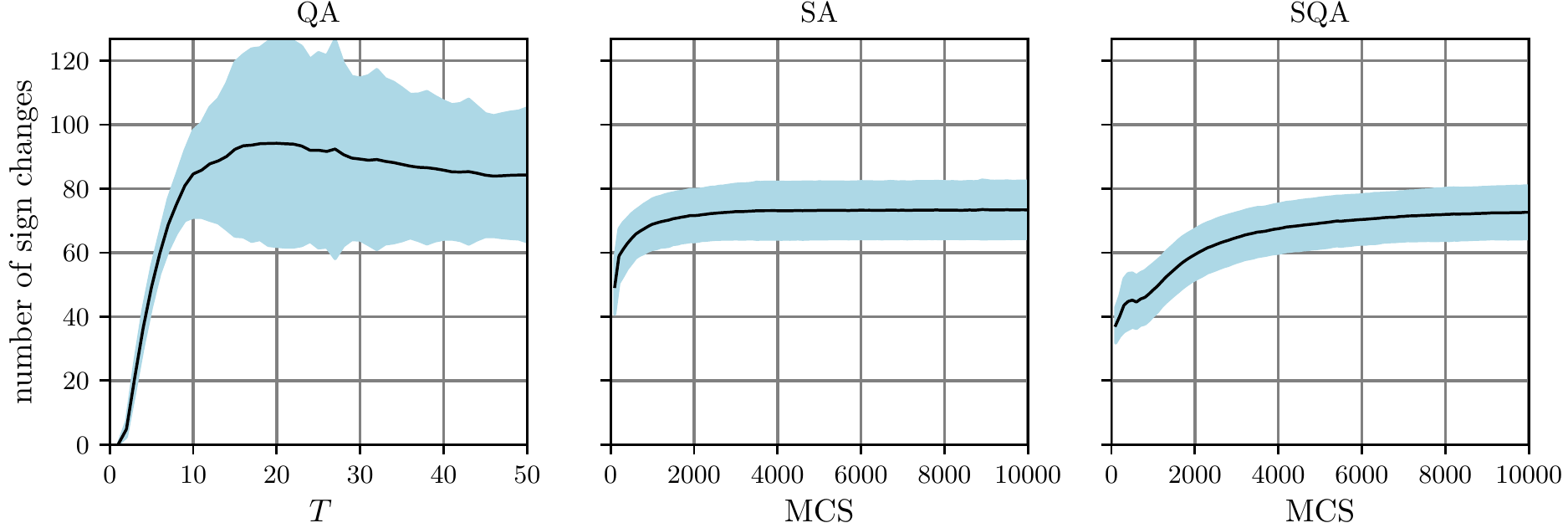}}
\caption{\label{fig: number of changes comparison}Number of sign changes in a full sweep averaged over all instances for $N=10$ (top) and $20$ (bottom). The filled-in region shows the standard deviation. For $N=10$ the average and deviations seem almost equivalent between the methods, while for larger $N$ a large peak with large deviation is developed at intermediate times for quantum annealing, while simulated and simulated quantum annealing remain very similar.}
\end{figure*}

While sign changes are not standard observables, they are still related to the underlying distribution in configuration space. Therefore, similarities here also indicate similarities in the dynamics of the distribution between the methods. A rough approach to quantify these similarities is to count the number of sign changes during the sweep. Since the number, as well as the location of the changes, depends on the annealing time, which one can not "map" between the different methods, comparisons of single sweeps can not be sensibly performed. Therefore, we need an overview of all timescales. One approach to get an overview is to look at the \emph{total number of sign changes} over a sweep as a function of the annealing time. The averages of this number over all instances for each system size is shown Fig.~\ref{fig: number of changes comparison}.
Here, the average number, as well as the standard deviation, is plotted over different annealing times for $N=10$ and $N=20$. For the smaller size, the averages and deviations seem almost equivalent. For increasing size though, a higher number at intermediate times along with larger deviations are observed in QA compared to the other methods, which are similar also for the larger size. Here one should note, that not only the timescales but also the time resolution is different for the methods, although the resolution in $s$, $\Delta s = 0.01$, is the same for all methods: for QA the resolution is $\Delta T = 1$ and for SA and SQA it is $\Delta \mathrm{MCS}=100$. Hence, some changes that appear and close very quickly might be missed, but it seems very unlikely that the much higher number for QA is solely due to resolution. These results indicate that the distribution dynamics is very similar between SA and SQA (for the model investigated), while there are differences to QA.

With this, we finish the qualitative discussion of correlations and will focus on the probability distribution for short times in the next section.


\section{\label{sec: annealing dist}Short-time expansion}
We have argued above, that the correlation functions are related to the underlying distribution in configuration space. In this section, we work out the short-time distribution explicitly and use the result to explain the initial sign structure. Furthermore, we show that the distribution corresponds to a high-temperature thermal distribution and hence QA can be related to thermal sampling at short times.
\subsection{\label{ssec: annealing dist - high temp}High-temperature expansion}
For the high-temperature expansion, we expand the Boltzmann density matrix
 \begin{equation*}
  \rho(\beta)=\erm^{-\beta \opnamed{H}{SK}}/Z, \; Z=\sum\limits_{z} \erm^{-\beta E_{z}},
 \end{equation*}
at $\beta=0$ with the goal of obtaining the occupations in the computational basis $\rho_{z}$ and the correlation functions $G_{ij}=\braket{\pauli{z}{i} \pauli{z}{j} }= \Tr\left[\rho \pauli{z}{i}\pauli{z}{j} \right]$. The index $z$ here indicates a state in the computational basis i.e. $z \equiv \ket{z_{1}, \hdots, z_{n}}$ with $z_{k}=\pm 1$, and $E_{z}$ the corresponding energy. The expansion up to second order is derived in detail in Appendix~\ref{app: high-temp exp}. The resulting density matrix is 
\begin{equation}
\label{eq: qa ht exp rho_z}
\rho_{z} \approx \frac{1}{\mathcal{D}}\left[1-\beta E_{z} + \frac{\beta^{2}}{2}\left(E_{z}^{2}-\frac{N(N-1)}{2}(1-E_{z}) \right) \right],
\end{equation}
where $\mathcal{D}$ is the Hilbert space dimension. Using this, the expression for the correlation functions can also be obtained to be (see Appendix for details)
\begin{equation}
\label{eq: qa ht corrs zz}
G_{nm} \approx -\beta J_{nm} + \frac{\beta^{2}}{2}\left(2(J^{2})_{nm} +\frac{N(N-1)}{2}J_{nm}\right).
\end{equation}
Note that the first term implies, that for high-temperature the correlation functions $G_{nm}$ are proportional to $-J_{nm}$, which is precisely the behavior we have seen in the short-time dynamics of the correlation functions in the earlier section. This suggests, that there could be a relation between the dynamical distribution to the high-temperature distribution. We will derive such a relation for quantum annealing in the next subsection.
\subsection{\label{ssec: annealing dist - qa}Quantum annealing}
In a short-time expansion, we approximate the propagator $U$ using the Dyson series up to second order; from which we then obtain the occupations $\rho_{z} \propto |\braket{z|U|+, x}|^2$. The detailed derivation is shown in Appendix~\ref{app: short-time exp} and leads to the result
\begin{equation}
\label{eq: qa short time dist}
\rho_{z}=|\braket{z|U(t)|+, x}|^2 \approx \frac{1}{\mathcal{D}}\left[1-\frac{T^{2}}{3}E_{z}+O(T^{4}) \right],
\end{equation}
wherein only even powers of $T$ appear, due to the different factors of $i$ in the Dyson series. It is remarkable that even for very short times the ground state is already the most probable state, although the absolute probability is presumably too small to be useful in applications.  Comparing the first terms to the high-temperature expansion from Eq.~\eqref{eq: qa ht exp rho_z}, we see that the first terms agree if we define $\beta_{\mathrm{QA}}=T^{2}/3$. This also implies that the second order term in the high-temperature expansion would have to be matched to the $T^{4}$ term if a matching order by order should be possible. We strongly doubt that it is, since the correction terms we get from the Dyson series already include expressions not seen in the high-temperature expansion. While it can not be ruled out at this point, that some are canceled by contributions from the fourth-order Dyson series, this seems exceedingly unlikely. This detail notwithstanding, it is clear from the numerical simulations, that the time-evolved state \emph{is not} close to a thermal state at all annealing times, therefore there have to be corrections at some order. These corrections can include powers of $N$ and $E_{z}$ and some expressions based on the bond matrix. Since both $N$ and $E_{z}$ grow with system size, albeit with different scaling, we expect that the corrections grow with system size and lead to a vanishing range of applicability of the expansion, as long as the value of $T$ is not scaled down with system size accordingly. Finally, we notice that the distribution in Eq.~\eqref{eq: qa short time dist} is normalized to the required order, since $\Tr[\opnamed{H}{SK}]=0$.

\begin{figure*}
\centering
\includegraphics[width=1.\linewidth]{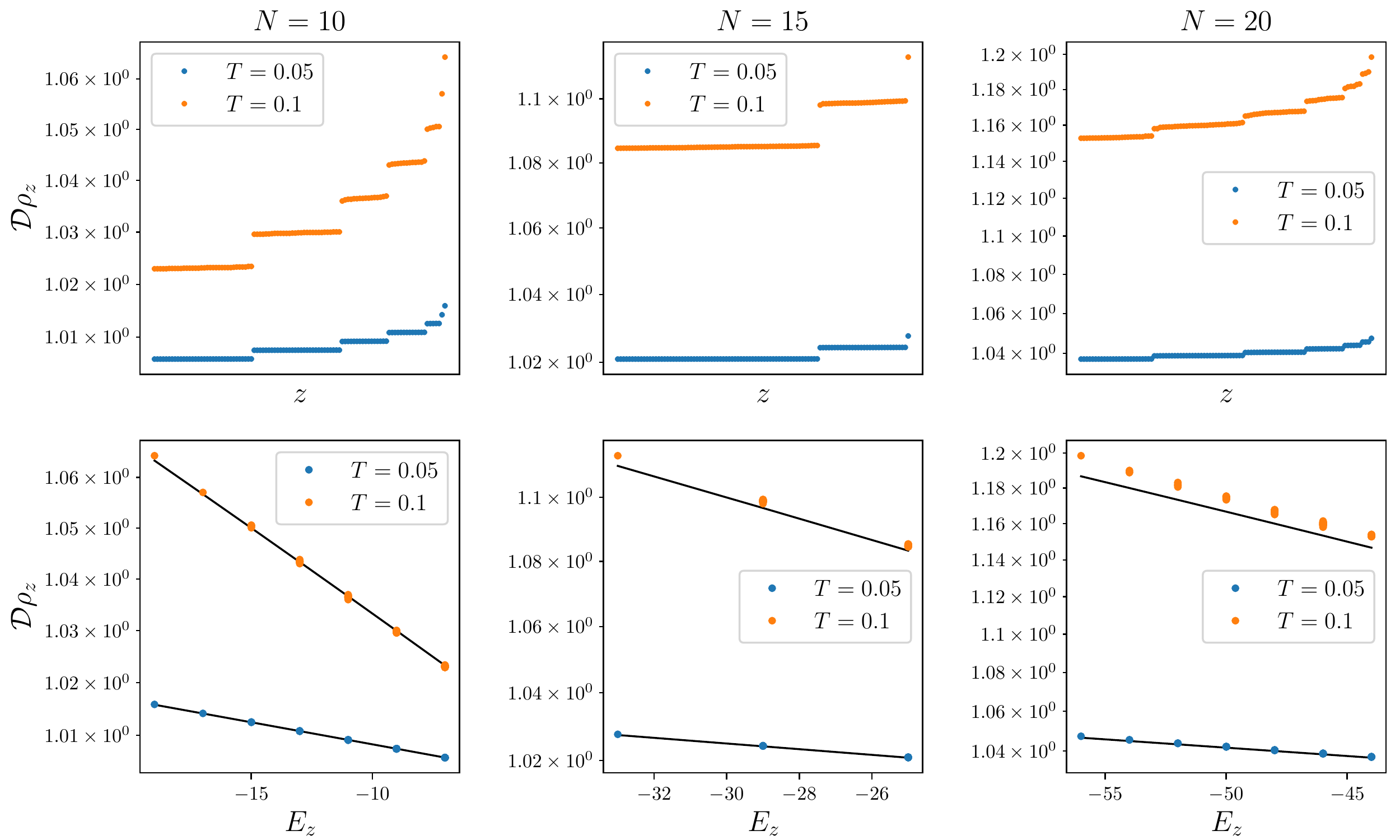} 
\caption{\label{fig: qa short dist} Scaled occupations of the $100$ most probable states in the computational basis at the end of the protocol for two annealing times $T=0.05$ and $T=0.1$. Three sizes, $N=10, 15, 20$, with one instance each are shown. In the upper plots, the occupations are ordered by their magnitudes. Plateaus arise, due to the high degeneracy of the energy levels, as discussed in Sec.~\ref{ssec: annealing model - sk}. For a thermal distribution, the plateaus would be exact for any temperature, therefore deviations seen at larger times indicate a non-thermal state and not just higher order thermal corrections. In the lower plots, the occupations are plotted as a function of the energy $E_{z}$. The lines indicate the high-temperature estimate with $\beta=T^{2}/3$. For the shorter time, these estimates agree well at all sizes, while there are clear deviations for the larger time. Furthermore, the occupations do not perfectly overlap for the larger time, again indicating that they are not a function of energy at this time.}
\end{figure*}

We test the calculations by directly probing the occupations in the computational basis $\rho_{z}$. We do this by computing and analyzing the $100$ most probable states in that basis. The results are depicted in Fig.~\ref{fig: qa short dist}. Here, we plot the occupations scaled by the Hilbert space dimension $\mathcal{D}\rho_{z}$ for two annealing times ($T=0.05$ and $T=0.1$). For each instance, the occupations are plotted twice: sorted by their magnitude in the upper plot and as a function of the energy $E_{z}$ on the lower plot. For a thermal distribution at \emph{any} temperature, we would expect the occupations to be pure functions of the energy. Due to the degeneracy of the energies, explained in Sec.~\ref{ssec: annealing model - sk}, this would lead to multiple "sharp" equidistant plateaus in the upper plots. Indeed, we observe plateaus, whose sharpness however decreases with annealing time. Already for $T=0.1$, deviations are seen, which only increase for even larger times (not shown in the figure). Therefore, already at this time the state is non-thermal and can not be described by a higher order thermal expansion. The same conclusion is confirmed by the lower plots. For a thermal state, all points would overlap and collapse to a single point at each energy, however, the points are separated for the larger time. The lines in the lower plot show the estimates based on the high-temperature expansion ($\beta = T^{2}/3$), with a good agreement seen at $T=0.05$ and some deviations at $T=0.1$. The deviations increase with system size, in agreement with the observations for the correlation functions. 

To finish the discussion, we would like to emphasize that this is a nontrivial result. Even the fact that the ground state is the most probable state at short times is not obvious (to us) \textit{a priori}, let alone the observation that the occupations are ordered in a thermal manner. To our best knowledge, the only other work investigating similar questions for short sweeps is~\cite{callison_energetic_2021}. While that work uses different methods, it seems that some of the results, for example an increase of the energy expectation value, are consistent with our calculations.

\subsection{\label{ssec: annealing dist - ca}Classical annealing}
\begin{figure*}
\centering
\includegraphics[width=1.\linewidth]{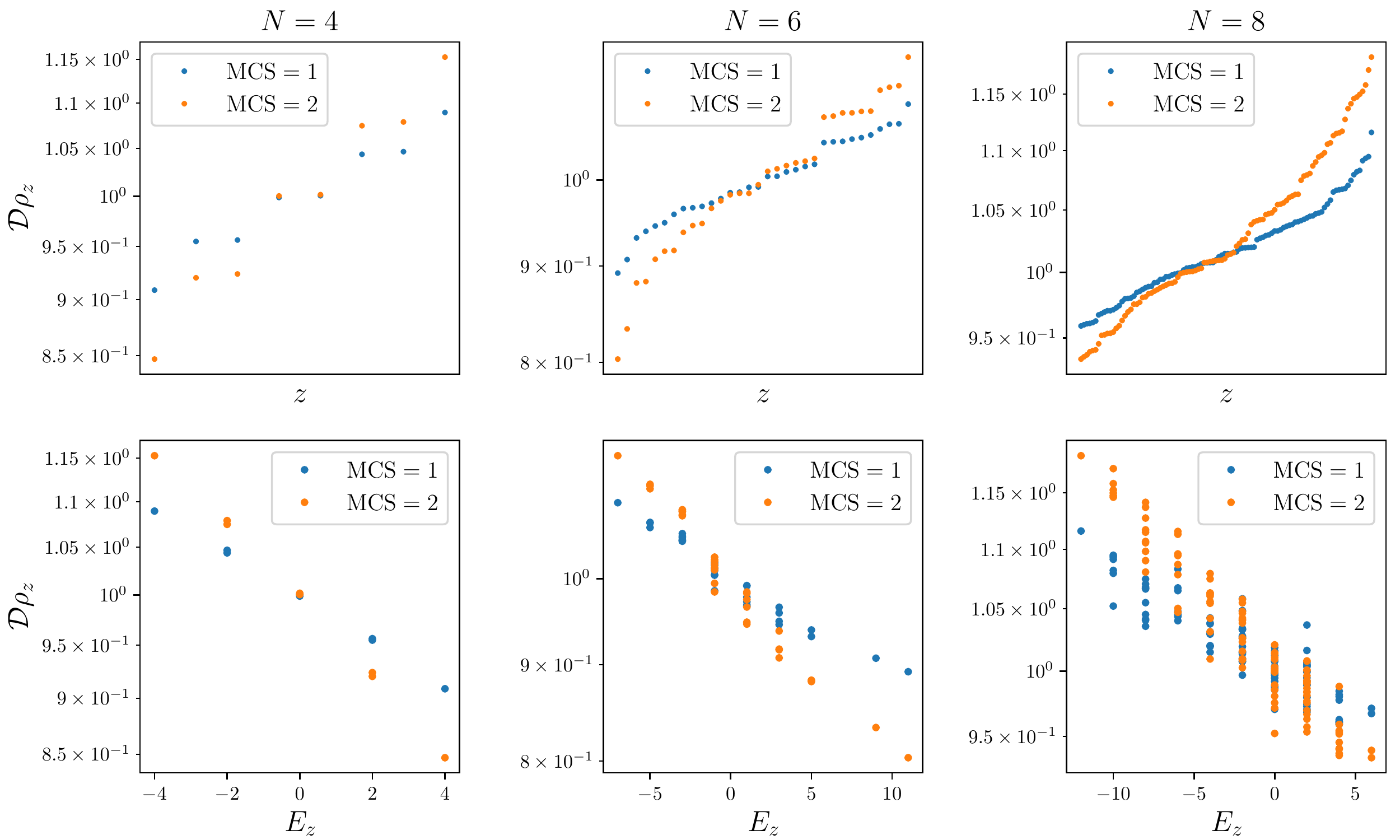} 
\caption{\label{fig: sqa short dist}Distribution of the most probable occupations in the computational basis after simulated quantum annealing. The data for three small sizes $N=4, 6, 8$ and the shortest possible annealing times $\mathrm{MCS}=1, 2$ are shown. The upper plots show the occupations sorted by their magnitude, while the lower ones show them as a function of $E_{z}$. For the smallest size, the results resemble the thermal distribution, but already at the next size, the corrections become significant.}
\end{figure*}
As for the fidelity, an analytical treatment of classical annealing is more involved than the one for quantum annealing. Given the similarity of the Schrödinger equation and the master equation, described in Sec.~\ref{sec: annealing methods}, one might have the idea to use a Dyson series analogously for the classical case. This requires a representation of the transition rate matrix, suitable for computing the integrals and the matrix elements. Representations using the Pauli matrices have been reported in the literature~\cite{wild_quantum_2021, wild_quantum_2021a}, but why we could use them to represent the transition matrix in our model, we did not find a way to integrate the corresponding expressions and determine appropriate matrix elements. Generally, it seems reasonable though that an algorithm like simulated annealing produces a distribution resembling a high-temperature distribution at short times since the relaxation and mixing times at those temperatures can also be small as noted in Sec.~\ref{ssec: annealing fidelity - classical annealing}. As before, we performed short-time simulations for simulated quantum annealing, which are shown in Fig.~\ref{fig: sqa short dist}. The layout is the same as in Fig.~\ref{fig: qa short dist}: the upper plots show the occupations ordered by magnitude and the lower as a function of $E_{z}$. However, even for a \emph{single} Monte Carlo step~\footnote{The step, in this case, corresponds to the middle of the parameter range $s=\frac{1}{2}$.}, a thermal distribution is only observed at $N=4$. Thus, a relationship between the annealing time and $\beta$ can not sensibly be extracted from our simulations. Again, the power law growth of correlation functions has origins different from the high-temperature expansion.
\section{\label{sec: annealing conclusion}Conclusion and outlook}
In this work, we have studied quantum and classical annealing of the Sherrington-Kirkpatrick model with various annealing times. Our comparative study with the same instances and a range of annealing times for each method is particularly suited to discern qualitative differences between those methods. While the long-time adiabatic behavior is relatively well studied, the behavior at faster annealing times is less well understood, but could be useful.

We identified simple and hard instances, which show striking differences in quantum annealing, highlighted by the appearance of a diabatic bump in the evolution of the fidelity for hard instances. Such strong differences were not observed for the classical methods. An open question is the accurate estimation of the time at which the diabatic bump appears. We have discussed the relevant factors but did not develop a quantitative theory. A good estimate may have important practical implications though, as at the bump we observe a relatively high fidelity, hence it could be advantageous to make multiple measurements around the bump time than to use fewer but larger annealing runs.

An analysis of two-point correlation functions indicated three dynamical regimes based on annealing times for all methods. We then used the number of sign changes as a qualitative measure of similarity, finding that the average number of sign changes is higher in quantum annealing than in the classical variants, which both show very similar numbers. Finally, we used a short-time expansion to show that quantum annealing produces a high-temperature thermal state at very short times. This also explains the qualitative short-time behavior of the correlation functions, although as was observed, the powers match only for very short times, while the qualitative behavior seems to persist longer.

\begin{acknowledgments}
We acknowledge support by the Austrian Science Fund FWF within the DK-ALM (W1259-N27). 

The computations and figures in this work have been obtained mostly using \textit{Python}~\cite{langtangen_primer_2009}, in particular with the (free and open) libraries \textit{Numba}~\cite{lam_numba_2015}, \textit{Numpy}~\cite{vanderwalt_numpy_2011, oliphant_guide_2015}, \textit{SciPy}~\cite{virtanen_scipy_2020} and \textit{Matplotlib}~\cite{hunter_matplotlib_2007},  and partially using \textit{Mathematica}~\cite{wolfram_mathematica_2020}. The data and code for this article is accessible in an open Zenodo archive~\cite{rakcheev_dataset_2022a}.
\end{acknowledgments}

\appendix
\section{\label{app: high-temp exp} High-temperature expansion}
The high-temperature expansion is derived starting from the Taylor series 
\begin{align*}
\rho = & \; \erm^{-\beta \opnamed{H}{SK}}/Z = \frac{\sum\limits_{n=0}^{\infty}\frac{(-\beta)^{n}}{n!}\opnamed{H}{SK}^{n}}{\sum\limits_{z}\sum\limits_{n=0}^{\infty}\frac{(-\beta)^{n}}{n!} E_{z}^{n}} \\
\approx & \;\frac{1}{\mathcal{D}}\left( 1-\opnamed{H}{SK}+\frac{\beta^{2}}{2}\opnamed{H}{SK}^{2}\right)\times\\ 
& \Bigg( 1+\beta \frac{\Tr\left[ \opnamed{H}{SK}\right]}{\mathcal{D}} + \beta^{2}\frac{\left(2\Tr\left[ \opnamed{H}{SK}\right]^{2}-\mathcal{D}\Tr\left[ \opnamed{H}{SK}^{2}\right] \right)}{2\mathcal{D}^{2}} \Bigg).
\end{align*}
We can simplify the expression, by evaluating the traces involved. Since the Pauli matrices are traceless, the sum over all configurations of any product, for example, $\pauli{z}{1}\pauli{z}{3}$, vanishes. For this reason,
 \begin{equation}
  \Tr\left[ \opnamed{H}{SK}\right]=0,
 \end{equation}
 since all indices involved are different. For the square term, however, we have a non-vanishing trace, since this also includes products with the same index and $(\pauli{z}{i})^{2}=I$, which is not traceless. The evaluation gives
\begin{align*}
\Tr\left[ H^{2}_{SK}\right] &= \sum\limits_{z} \sum\limits_{i_{1}>j_{1}, i_{2}>j_{2}}J_{i_{1}j_{1}} J_{i_{2}j_{2}} \pauli{z}{i_{1}}\pauli{z}{j_{1}} \pauli{z}{i_{2}}\pauli{z}{j_{2}}\\ 
&= \sum\limits_{z} \sum\limits_{i_{1}>j_{1}, i_{2}>j_{2}}J_{i_{1}j_{1}} J_{i_{2}j_{2}} \delta_{(i_{1}, j_{1}),(i_{2}, j_{2})} \\ 
&=  \sum\limits_{z} \sum\limits_{i_{1}>j_{1}}J^{2}_{i_{1}j_{1}}=\mathcal{D}\frac{N(N-1)}{2},
\end{align*}
since the only way to remove the Pauli operators, and hence make the trace non-vanishing, is to choose equal indices. Since the indices are ordered, only one assignment $i_{1}=i_{2}, \; j_{1}=j_{2}$ is possible. The occupations to second order in $\beta$, are therefore
\begin{equation*}
\rho_{z} \approx \frac{1}{\mathcal{D}}\left[1-\beta E_{z} + \frac{\beta^{2}}{2}\left(E_{z}^{2}-\frac{N(N-1)}{2}(1-E_{z}) \right) \right].
\end{equation*}

Let us now evaluate the necessary expressions to compute the correlation functions $G_{nm}$ for $n \neq m$ 
\begin{align*}
\Tr\left[ \pauli{z}{n}\pauli{z}{m} \opnamed{H}{SK} \right] &= \sum\limits_{z} \sum\limits_{i > j} J_{ij} \pauli{z}{n}\pauli{z}{m}\pauli{z}{i}\pauli{z}{j} = \sum\limits_{z} J_{nm} = \mathcal{D} J_{nm}.
\end{align*}
For the next terms, we get
\begin{align*}
&\Tr\left[ \pauli{z}{n}\pauli{z}{m}  H^{2}_{SK} \right] = \frac{1}{4}\sum\limits_{z} \sum\limits_{i \neq j}\sum\limits_{k \neq l} J_{ij}J_{kl} \pauli{z}{n}\pauli{z}{m}\pauli{z}{i}\pauli{z}{j}\pauli{z}{k}\pauli{z}{k} \\
&= \frac{\mathcal{D}}{4} \sum\limits_{i \neq j}\sum\limits_{k \neq l} J_{ij}J_{kl}\Big(\delta_{nj}\delta_{jk}\delta_{km}+\delta_{nj}\delta_{jl}\delta_{lm}+\delta_{mj}\delta_{jk}\delta_{kn}\\
&\;\;\;\;\; +\delta_{mj}\delta_{jl}\delta_{ln} + \delta_{in}\delta_{ik}\delta_{km}+\delta_{in}\delta_{il}\delta_{lm}\\
&\;\;\;\;\;+\delta_{im}\delta_{ik}\delta_{kn}+\delta_{im}\delta_{il}\delta_{ln}\Big)\\
&= 2\mathcal{D}\sum\limits_{n \neq j \neq m}J_{nj}J_{jm} = 2\mathcal{D}\sum\limits_{j}J_{nj}J_{jm} =  2\mathcal{D} (J^{2})_{nm},
\end{align*}
where the last expression denotes the matrix elements of the squared bond matrix. In total therefore we get
\begin{equation*}
    G_{nm} \approx -\beta J_{nm} + \frac{\beta^{2}}{2}\left(2(J^{2})_{nm} +\frac{N(N-1)}{2}J_{nm}\right),
\end{equation*}
where the terms of $\rho_{z}$ without factors of $E_{z}$ do not contribute, due to the aforementioned tracelessness of Pauli matrices.
\section{\label{app: short-time exp} Short-time expansion}
For the short-time expansion, we will start from Dyson series, since this combines the expansions of dynamics with expanding into powers of operators. If we used the Magnus expansion instead, we would need to evaluate the matrix elements of exponentials of operators, rather than powers, which is generally not possible. The series up to the second order is
\begin{align*}
U(t) \approx &\; 1-\irm \int\limits_{0}^{t} \drm t_{1} \;  H(t_{1}/T)\\
&+(-\irm)^{2}\int\limits_{0}^{t} \drm t_{1}\int\limits_{0}^{t_{1}}\drm t_{2}\; H(t_{1}/T)H(t_{2}/T)
\end{align*}
Evaluation of these terms gives
\begin{align*}
&(-\irm)T\int\limits_{0}^{s}\drm s_{1} \; H(s_{1})= -\irm\frac{sT}{2}\Bigg[s\opnamed{H}{fin} + (2-s)\opnamed{H}{ini} \Bigg]\\
&(-\irm)^{2}T^{2}\int\limits_{0}^{s}\drm s_{1}\int\limits_{0}^{s_{1}}\drm s_{2} \; H(s_{1})H(s_{2})= -\frac{(sT)^{2}}{8}\Bigg[s^{2}\opnamed{H}{fin}^{2}  \\
&+ (2-s)^{2}\opnamed{H}{ini}^{2} + \frac{s(8-3s)}{3}\opnamed{H}{fin}\opnamed{H}{ini} + \frac{s(4-3s)}{3}\opnamed{H}{ini}\opnamed{H}{fin} \Bigg],
\end{align*}
where with the use of $\mathrm{ini}$ and $\mathrm{fin}$, we emphasize that this only depends on the schedule and not on the particular Hamiltonians at this point. We see that different powers of $s$ and $T$ can be mixed, so the ordering by powers of $sT$ will only be useful if $T$ is the dominant factor. In the following, we focus on the full sweep ($s=1$), where the expansion becomes an expansion in $T$. 

In our setup, $\opnamed{H}{ini}=\opnamed{H}{x}$ and $\opnamed{H}{fin}=\opnamed{H}{SK}$ and we need to evaluate the matrix elements of the operators in the expansion, between the $x$-polarized state and a product state in the computational basis $\braket{z|O|+, x}$. For the operators appearing in the expansion, these are
\begin{align*}
\braket{z|\opnamed{H}{SK}^{n}|+, x} &= E^{n}_{z}/\sqrt{\mathcal{D}}\\
\braket{z|\opnamed{H}{x}^{n}|+, x} &= (-N)^{n}/\sqrt{\mathcal{D}} \\
\braket{z|\opnamed{H}{SK}^{n}\opnamed{H}{x}^{m}|+, x} &= (-N)^{m}E^{n}_{z}/\sqrt{\mathcal{D}}\\
\braket{z|\opnamed{H}{x}\opnamed{H}{SK}|+, x} &=-(N-4)E_{z}/\sqrt{\mathcal{D}}.
\end{align*}
The first three can be derived simply by using the eigenvalues 
\begin{equation*}
\opnamed{H}{x}\ket{+, x}=-N\ket{+, x}, \; \opnamed{H}{SK}\ket{z}=E_{z}\ket{z},
\end{equation*}
and the decomposition of the $x$-polarized state 
\begin{equation*}
\braket{z|+, x} = \frac{1}{\sqrt{\mathcal{D}}},\; \forall z.
\end{equation*} 
For the last expression, some more steps are needed after inserting the decomposition
\begin{equation*}
 \braket{z|\opnamed{H}{x}\opnamed{H}{SK}|+, x} = \frac{1}{\sqrt{\mathcal{D}}}\sum\limits_{\tilde{z}}E_{\tilde{z}}\braket{z|\opnamed{H}{x}|\tilde{z}}.
\end{equation*}
Since the transverse field has a matrix element of $-1$ between states differing by a single spin-flip and vanishes otherwise, the sum includes the energies from all states differing by a single spin-flip. Writing the excitation energy from a flip of spin $n$ as $(\Delta E)_{n}=-z_{n}\sum\limits_{m}J_{nm}z_{m}$, we can evaluate the sum
\begin{align*}
&\frac{1}{\sqrt{\mathcal{D}}}\sum\limits_{\tilde{z}}E_{\tilde{z}}\braket{z|\opnamed{H}{x}|\tilde{z}} = -\sum\limits_{n}\left(E_{z}-z_{n}\sum\limits_{m}J_{nm}z_{m}\right)\\
=&-NE_{z}+\sum\limits_{n,m}J_{nm}z_{n}z_{m} = -(N-4)E_{z},
\end{align*}
where in the last line we used the definition of the energy $E_{z}=\frac{1}{2}\sum\limits_{n,m}J_{nm}z_{n}z_{m}$. Substituting these results into $|\braket{z|U(t)|+, x}|^2$, we obtain the density matrix from Eq.~\eqref{eq: qa short time dist}.
\bibliography{bib_qa}
\end{document}